\documentclass[]{aastex701}

\usepackage{verbatim}

\begin{document}

\title{Unveiling Hidden X-ray Broad Line Region in Centaurus A: Broadband Analysis with XRISM and NuSTAR}

\author[orcid=0009-0000-9577-8701,sname='Nakatani']{Yuya Nakatani}
\affiliation{Department of Astronomy, Kyoto University, Kitashirakawa-Oiwake-cho, Sakyo-ku, Kyoto 606-8502, Japan}
\email[show]{nakatani@kusastro.kyoto-u.ac.jp}

\author[orcid=0000-0001-7821-6715,sname='Yoshihiro Ueda']{Yoshihiro Ueda}
\affiliation{Department of Astronomy, Kyoto University, Kitashirakawa-Oiwake-cho, Sakyo-ku, Kyoto 606-8502, Japan}
\email[show]{ueda@kusastro.kyoto-u.ac.jp}

\author[orcid=0000-0002-5924-4822,sname='David Bogensberger']{David Bogensberger}
\affiliation{Universit\'e Paris-Saclay, Universit\'e Paris Cit\'e, CEA, CNRS, AIM, 91191 Gif-sur-Yvette, France}
\affiliation{Department of Astronomy, The University of Michigan, 1085 South University Avenue, Ann Arbor, Michigan, 48103, USA}
\email{}

\author[orcid=0000-0002-6960-9274,sname='Taishu Kayanoki']{Taishu Kayanoki}
\affiliation{Department of Physics, Graduate School of Advanced Science and Engineering, Hiroshima University, Hiroshima 739-8526, Japan}
\affiliation{Faculty of Engineering, University of Miyazaki, Miyazaki 889-2192, Japan}
\email{}

\author[orcid=0000-0002-0921-8837,sname='Yasushi Fukazawa']{Yasushi Fukazawa}
\affiliation{Department of Physics, Graduate School of Advanced Science and Engineering, Hiroshima University, Hiroshima 739-8526, Japan}
\email{}

\author[orcid=0009-0006-4968-7108,sname='Luigi Gallo']{Luigi Gallo}
\affiliation{Saint Mary’s University, Department of Astronomy and Physics, 923 Robie St, Halifax, Nova Scotia B3H 3C3, Canada}
\email{}

\author[orcid=0000-0002-7962-5446,sname='Richard Mushotzky']{Richard Mushotzky}
\affiliation{Department of Astronomy, University of Maryland, College Park, MD 20742, USA}
\affiliation{Joint Space-Science Institute, University of Maryland, College Park, MD 20742, USA}
\email{}

\author[sname='Tahir Yaqoob']{Tahir Yaqoob}
\affiliation{National Aeronautics and Space Administration, Goddard Space Flight Center, Greenbelt, MD 20771, USA}
\affiliation{Center for Research and Exploration in Space Science and Technology, NASA/GSFC (CRESST II), Greenbelt, MD 20771, USA}
\affiliation{Center for Space Science and Technology, University of Maryland, Baltimore County (UMBC), 1000 Hilltop Circle, Baltimore, MD 21250, USA}
\email{}

\author[orcid=0000-0001-8055-7113,sname='Kotaro Fukushima']{Kotaro Fukushima}
\affiliation{Department of Physics, Tokyo University of Science, 1-3 Kagurazaka, Shinjuku-ku, Tokyo 162-8601, Japan}
\email{}

\author[orcid=0000-0002-5488-1961,sname='Bert Vander Meulen']{Bert Vander Meulen}
\affiliation{European Space Agency (ESA), European Space Research and Technology Centre (ESTEC), Keplerlaan 1, 2201 AZ Noordwijk, The Netherlands}
\affiliation{Department of Physics and Astronomy, Universiteit Gent, Proeftuinstraat 86 N3, B-9000 Ghent, Belgium}
\email{}

\author[orcid=0000-0001-7557-9713,sname='Junjie Mao']{Junjie Mao}
\affiliation{Department of Astronomy, Tsinghua University, Beijing 100084, China}
\affiliation{Department of Physics, Graduate School of Advanced Science and Engineering, Hiroshima University, Hiroshima 739-8526, Japan}
\email{}

\author[orcid=0009-0006-4377-4219,sname='Kanta Fujiwara']{Kanta Fujiwara}
\affiliation{Department of Astronomy, Kyoto University, Kitashirakawa-Oiwake-cho, Sakyo-ku, Kyoto 606-8502, Japan}
\email{}

\author[orcid=0000-0003-4235-5304,sname='Kouichi Hagino']{Kouichi Hagino}
\affiliation{Department of Physics, The University of Tokyo, 7-3-1 Hongo, Bunkyo-ku, Tokyo 113-0033, Japan}
\email{}

\author[orcid=0000-0002-2304-4773,sname='Toshiya Iwata']{Toshiya Iwata}
\affiliation{Department of Physics, The University of Tokyo, 7-3-1 Hongo, Bunkyo-ku, Tokyo 113-0033, Japan}
\affiliation{RIKEN Pioneering Research Institute, 2-1 Hirosawa, Wako, Saitama 351-0198, Japan}
\email{}

\author[orcid=0000-0003-2869-7682,sname='Jon M. Miller']{Jon M. Miller}
\affiliation{Department of Astronomy, The University of Michigan, 1085 South University Avenue, Ann Arbor, Michigan, 48103, USA}
\email{}

\author[orcid=0000-0003-2161-0361,sname='Misaki Mizumoto']{Misaki Mizumoto}
\affiliation{Science Education Research Unit, University of Teacher Education Fukuoka, Akama-bunkyo-machi, Munakata, Fukuoka 811-4192, Japan}
\email{}

\author[orcid=0000-0001-6020-517X,sname='Hirofumi Noda']{Hirofumi Noda}
\affiliation{Astronomical Institute, Tohoku University, 6-3 Aramakiazaaoba, Aoba-ku, Sendai, Miyagi 980-8578, Japan}
\email{}

\author[orcid=0000-0002-5097-1257,sname='Makoto Tashiro']{Makoto Tashiro}
\affiliation{Department of Physics, Saitama University, Saitama 338-8570, Japan}
\affiliation{Institute of Space and Astronautical Science (ISAS), Japan Aerospace Exploration Agency (JAXA), Kanagawa 252-5210, Japan}
\email{}

\author[orcid=0009-0008-3206-235X,sname='Misaki Urata']{Misaki Urata}
\affiliation{Department of Physics, Graduate School of Advanced Science and Engineering, Hiroshima University, Hiroshima 739-8526, Japan}
\email{}

\begin{abstract}
  We present a broadband X-ray spectral analysis of the radio galaxy Centaurus~A, utilizing simultaneous \textit{XRISM} (Resolve and Xtend) and \textit{NuSTAR} observations.
  Unlike the previous phenomenological analysis of the Fe K$\alpha$ line (Bogensberger et al. 2025), we employ physical reflection models developed with the radiative transfer code SKIRT.
  They self-consistently predict the continuum and fluorescent lines from a clumpy torus and an X-ray broad line region (xBLR).
  The Fe K$\alpha$ complex is well reproduced with a three velocity component model.
  We interpret this as two narrow torus components (FWHM $= 310\pm50$ and $1900\pm400$~km~s$^{-1}$) and an xBLR component (FWHM $= 6000_{-800}^{+1100}$~km~s$^{-1}$).
  The inner torus radius of 0.13~pc is 10 times the dust sublimation radius (0.013~pc), 
  implying that the torus in a low-luminosity AGN is not determined by dust-related physics.
  The xBLR has a substantial covering factor of 0.11--0.36, a hydrogen column density ($5 \times 10^{23}$~cm$^{-2}$), a radius of 0.014~pc from the SMBH, and a mass of $>0.9 M_\odot$.
  Its mass supply rate is comparable to the accretion rate inferred from the bolometric luminosity, implying efficient xBLR formation even at low Eddington ratios.
  The xBLR line width also provides a novel estimate of the black-hole mass, $\log{(M_{\rm BH}/M_{\odot})} = 7.00\pm0.34$, via the optical BLR mass-scaling relation to the Fe~K$\alpha$.
  This agrees with dynamical measurements within a factor of $\sim5$, consistent with the scatter of the optical relation.
  We also infer the abundance ratios of $A_{\rm Ca}/A_{\rm Fe} = 1.1_{-0.1}^{+0.2}$ and $A_{\rm Ni}/A_{\rm Fe} = 1.4_{-0.2}^{+0.1}$ solar.
\end{abstract}

\keywords{\uat{Galaxies}{573} --- \uat{High Energy astrophysics}{739}}

\section{Introduction}\label{sec:intro}

A tight correlation exists between the mass of the central supermassive black hole (SMBH) and the properties of the host galaxy's bulge component, suggesting a fundamental link between these two components (e.g. \citealt{Magorrian+98, Kormendy_Ho_2013}).
This implies a co-evolutionary scenario driven by feeding and feedback processes, bridging spatial scales that differ by more than nine orders of magnitude (see \citealt{Fabian_2012} for a review).
In the context of feedback from the SMBH to the host galaxy, the Active Galactic Nucleus (AGN) phase, during which the SMBH actively accretes mass, plays a crucial role.
Observationally, AGNs are classified into two types: type-1 AGNs, which exhibit broad emission lines with a Full Width at Half Maximum (FWHM) of typically a few thousand to ten thousand km~s$^{-1}$ in the UV, optical, or infrared bands
and type-2 AGNs, which lack these broad features.
The unified model, in which the difference between type-1 and type-2 AGNs is attributed to the viewing angle of the observer relative to a dusty torus surrounding the central engine, has been successful in explaining this dichotomy (\citealt{Antonucci_1993}).
However, a complete view of the matter distribution and kinematics of the central engine (accretion disk, broad line region (BLR), and torus) has remained elusive due to observational biases.

AGN feedback mainly operates through jets, winds, and radiation (e.g. \citealt{King_Pounds_2015}).
Radio galaxies (RGs), characterized by powerful relativistic jets, are essential for studying AGN feedback by jets.
These jets propagate at relativistic speeds and interact with the interstellar medium (ISM), thereby regulating star formation in the host galaxy.
Although jet launching and feedback efficiency are thought to be linked to the central accretion flow (\citealt{Blandford_Znajek_1977}),
the detailed mechanisms and their connection to the subparsec-scale nuclear structure remain poorly understood.

X-ray observations are essential for probing the nuclear structure of
heavily obscured AGNs.  X-rays penetrate the dense absorbing material
(the column density along the line of sight with $\log{N_{\rm H}^{\rm
    los}/{\rm cm}^{-2}} > 22$) to reveal the nuclear structure and the
circumnuclear gas and dust across diverse thermal phases.
Historically, detailed spectral decomposition has been hindered by the limited energy resolution of X-ray instruments 
(e.g., $E/\Delta E \sim 50$ for \textit{XMM-Newton}/EPIC-pn, and $E/\Delta E \sim 100\textendash200$ for \textit{Chandra}/HETG at 6~keV). 
\textit{XRISM} (\citealt{Tashiro+21, Tashiro+24}), launched in 2023 by JAXA, 
carries a microcalorimeter (Resolve) with superior energy resolution (FWHM $\sim$4--5~eV, or $E/\Delta E \sim$ 1200--1500 at 6~keV).
Resolve allows us to spectroscopically resolve the velocity fields and physical conditions of the gas on spatial scales that are not accessible to imaging.
For example, observations of the Circinus Galaxy (type-2) successfully constrained the chemical abundances in the torus, revealing its material origin (\citealt{XRISM_Circinus2026}). 
Furthermore, in the complex type-1.5 AGN NGC~4151, which exhibits broad optical lines but significant X-ray absorption,
\textit{XRISM} identified a distinct X-ray broad line region (xBLR), providing new constraints on the nuclear geometry (\citealt{XRISM_NGC4151_2024}).
Subsequent \textit{XRISM} studies of NGC~4151 have revealed stratified disk winds and the reverberation of the narrow Fe K$\alpha$ line (e.g., \citealt{Xiang+25, Miller+26}).

Our target, Centaurus~A (hereafter Cen~A), is one of the nearest RGs located at a distance of $3.8\pm0.1$~Mpc (\citealt{Harris+10}), 
classified as a heavily obscured ($\log{N_{\rm H}^{\rm los}/{\rm cm}^{-2}} = 23\pm 0.2$) type-2 AGN (\citealt{Beckmann+11}). 
Due to its strong X-ray and radio flux, Cen~A has been extensively studied as a key object for understanding AGN physics. 
The mass of the central SMBH ($M_{\rm BH}$) has been measured with several kinematic methods.
Using VLT/SINFONI, \citet{Cappellari+09} and \citet{Neumayer+07} derived
$M_{\rm BH} = (5.5\pm3.0)\times10^{7}\,M_{\odot}$ 
and $M_{\rm BH} = 4.5^{+1.7}_{-1.0}\times10^{7}\,M_{\odot}$ (errors are $3\sigma$)
from modeling of stellar kinematics and of molecular gas, respectively, for a distance of
3.5~Mpc\footnote{Since the masses scale linearly with the assumed distance (\citealt{Cappellari+09}), they would increase by $0.04$~dex for a distance of 3.8~Mpc \citep{Harris+10}.}.
Since the two measurements based on different tracers agree with each other,  
we adopt $\log{(M_{\rm BH}/M_{\odot})} = 7.7$ (their logarithmic mean) in our paper, 
which is also consistent with the recent result by \citet{Gnerucci+11}.
We note earlier kinematic studies reported somewhat larger masses 
($2.4^{+0.3}_{-0.2}\times10^{8}\,M_{\odot}$ based on stellar kinematics, \citet{Silge+05};
$(1.1\pm0.1)\times10^{8}\,M_{\odot}$ from [S~III], Pa$\beta$ and [Fe~II] rotation curves, \citet{Marconi+06};
$8.25^{+2.25}_{-4.25}\times10^{7}\,M_{\odot}$ from the two-dimensional Pa$\beta$ velocity field modeling, \citet{Krajnovic+07}), which
range from $\log{(M_{\rm BH}/M_{\odot})} = 7.92$ to $8.38$ for a distance of 3.5 Mpc.
The adopted mass ($\log{(M_{\rm BH}/M_{\odot})} = 7.7$) corresponds to a gravitational radius $r_{\rm g} = GM_{\rm BH}/c^2$ of about $7.4\times 10^{12}$~cm.
The bolometric luminosity is estimated to be $L_{\rm bol} \sim 1.7 \times 10^{43}$~erg~s$^{-1}$ (\citealt{Koss+22}),
resulting in a low Eddington ratio of $\lambda_{\rm Edd} = L_{\rm bol}/ 1.5 \times10^{38} (M_{\rm BH}/M_{\odot}) \sim 0.002$.

The classification of Cen~A as an RG does not limit the generality of the results presented here.
\citet{Nakatani+23} analyzed the broad-band \textit{Suzaku} and \textit{NuSTAR} spectra of seven obscured narrow-line radio galaxies, including Cen~A (\citealt{Ogawa+21}).
They used the XCLUMPY model (\citealt{Tanimoto+19}), which describes the reflection from a clumpy torus based on Monte Carlo simulations.
They found that the torus covering fractions of these radio galaxies are similar to those of radio-quiet AGNs at the same Eddington ratios.
This indicates that the geometric structure of the torus is primarily determined by the Eddington ratio, regardless of the presence of a radio jet.
Cen~A, despite being an RG, can thus be regarded as a representative obscured AGN.
Despite extensive searches in optical, infrared, and polarized light, no broad emission lines have been found (\citealt{Alexander+99, Marin+25, Goold+26}).
For example, \citet{Alexander+99} obtained near-infrared ($K$-band) spectropolarimetry of the nuclear region and searched for broad Br$\gamma$. 
They detected none in either total or polarized flux.
Their result was inconclusive, in the sense that a completely hidden BLR could not be distinguished from the intrinsic absence of one.
More recently, \textit{JWST} observations of the nucleus have also shown no broad Pa$\alpha$ component (\citealt{Goold+26}).
The visual extinction toward the near-infrared nucleus is estimated to be $A_V \approx$ 22--55~mag 
from the 9.7~$\mu$m silicate absorption feature (\citealt{Grasdalen_Joyce_1976, Turner+92}).
Whether the BLR in Cen~A is deeply hidden or intrinsically absent has thus remained an open question.
On larger scales, ALMA has resolved the circumnuclear molecular gas of Cen~A.
\citet{Espada+17} imaged the circumnuclear disk (200~pc) in CO(3--2) at a resolution of $\sim$5~pc.
They found an inner molecular ring with a deprojected radius of 50--100~pc, along with filamentary structures interpreted as nuclear shocks.
These structures may facilitate accretion onto the SMBH.
On sub-parsec scales, mid-infrared interferometry with VLTI/MIDI resolved a dusty disk with a diameter of $\sim$0.6~pc (\citealt{Meisenheimer+07}).
However, none of these observations constrain the kinematics of the obscuring and line-emitting material on sub-parsec scales.
Such kinematic information is accessible only through high-resolution X-ray spectroscopy.
In the X-ray band, Cen~A has been observed by various X-ray satellites, revealing its detailed spectral characteristics.
Its X-ray spectrum is characterized by strong absorption and a prominent Fe K$\alpha$ fluorescence line (intensity of the 6.4~keV line $I_{64} \sim (24 \pm 1)\times10^{-5}$~photons~cm$^{-2}$~s$^{-1}$; \citealt{Fukazawa+11}).
The reflection features of Cen~A have been described by several torus models, including MYTorus (\citealt{Murphy_Yaqoob_2009, Fuerst+16}), BNTorus (\citealt{Brightman_Nandra_2011, Fuerst+16}), and XCLUMPY (\citealt{Tanimoto+19, Ogawa+21}).
These studies suggest that the obscuring torus is characterized by an angular width of $50\textendash60$~deg (corresponding to a covering factor of $\sim 0.5$) and an inclination angle of $50\textendash75$~deg.
We note that previous estimates of the system's inclination range from edge-on ($50\textendash80$~deg; \citealt{Fuerst+16, Ogawa+21}) to face-on ($<45$~deg; \citealt{Hardcastle+03, Mueller+14, Bogensberger+24}).

Recent high-resolution spectroscopy with \textit{XRISM} has revealed a
broad iron line feature in addition to the dominant narrow Fe K$\alpha$ line core, suggesting the presence of a hidden BLR in
Cen~A (\citealt{Bogensberger+25}). However, a definitive physical
interpretation is challenged by the spectral degeneracy between the
line profile and the underlying reflection continuum.  To overcome
this, we perform a joint analysis using simultaneous \textit{NuSTAR}
data, which covers the hard band ($>10$~keV) essential for
constraining the Compton hump. We employ physical reflection models
based on the SKIRT radiative transfer code (\citealt{Baes+03, VanderMeulen+23})
to distinguish between BLR and torus contributions. In the latest
SKIRT code, the realistic line profiles measured in ground experiments
(\citealt{Hoelzer+97}) have been incorporated (\citealt{VanderMeulen+24}), which are known to well
reproduce the neutral Fe K$\alpha$ profile in a type-2 AGN
(\citealt{XRISM_Circinus2026}).

The organization of this paper is as follows. In Section~\ref{sec:obsdata} we describe
the observations of \textit{XRISM} and \textit{NuSTAR} and data reduction. Section~\ref{sec:xrayanalysis}
presents the details of our spectral analysis and results. The nuclear
structure of Cen~A is discussed in Section~\ref{sec:discussion}, including the physical
origin of the xBLR. Conclusions are summarized in Section~\ref{sec:conclusion}. Throughout
the paper, errors in spectral parameters are quoted at the 90\%
confidence limit for a single parameter. The solar abundances refer to
the protosolar table of \citet{Lodders+09}.

\section{Observation and Data Reduction}\label{sec:obsdata}
The X-ray data of Cen~A used in this study were obtained from two observatories, 
\textit{XRISM} and \textit{NuSTAR}, which provide complementary capabilities in terms of energy coverage and spectral resolution.
\textit{XRISM}/Resolve offers high energy resolution in the soft X-ray band (2$\textendash$12~keV).
\textit{XRISM}/Xtend is an X-ray CCD camera with a larger effective area in the soft X-ray band (0.3$\textendash$12~keV).
\textit{NuSTAR} provides broad energy coverage up to 79~keV with moderate energy resolution.
The details of the observations and data reduction processes for each instrument are described in the following subsections.
The observation logs are summarized in Table~\ref{tab:obslog}.

\begin{table*}
  \caption{Summary of observations.\label{tab:obslog}
    Exposures are based on the good time intervals of Resolve for \textit{XRISM} and those of FPMA for \textit{NuSTAR}.
    \newline
  }
  \begin{tabular}{lcccccc}
    \hline
    Observatory & ObsID & Start Date (UT) & End Date (UT) & Exposure (ks) & Energy Range (keV)\\
    \hline
    \textit{XRISM}  & 300019010   & 2024 Aug 04 19:17:04 & 2024 Aug 09 16:24:04 & 225 & 3--10 \\
    \textit{NuSTAR} & 60901034002 & 2024 Aug 05 07:01:09 & 2024 Aug 05 19:26:09 & 25 & 8--75 \\
    & 60901034004 & 2024 Aug 09 02:16:09 & 2024 Aug 10 19:21:09 & 77 & 8--75 \\
    \hline
  \end{tabular}
\end{table*} 

\subsection{XRISM}\label{subsec:xrism}
\textit{XRISM} carries two focal plane instruments, each coupled with an
X-ray Mirror Assembly (\citealt{Hayashi+24}): the high-energy-resolution X-ray
microcalorimeter Resolve (\citealt{Ishisaki+25, Kelley+25}) and the X-ray CCD camera Xtend (\citealt{Noda+25}).
\textit{XRISM} observed Cen~A in Aug 2024, for 225~ks (Table~\ref{tab:obslog}).
We use both the Resolve and Xtend data in this work.
The Resolve data provide the line profiles at high energy resolution.
The Xtend data provide an independent measurement of the continuum shape in the soft X-ray band.
Cross-calibration uncertainties among the instruments are taken into account by free multiplicative constants (Section~\ref{subsec:spectralmodel}).
The Resolve and Xtend data were reduced following the \textit{XRISM} Quick-Start Guide version 3.2
\footnote{\url{https://heasarc.gsfc.nasa.gov/docs/xrism/analysis/quickstart/xrism_quick_start_guide_v3p2_251204a.pdf}}.
The same observation was first analyzed by \citet{Bogensberger+25}.
We utilized the \textit{XRISM} calibration database (CalDB) version 20260315 for both Resolve and Xtend.
For this observation, the gain uncertainty is estimated to be less than 1~eV
at 6.4~keV.  The Resolve spectrum was binned according to the optimal binning
scheme by \citet{Kaastra_Bleeker_2016} through the \texttt{ftgrouppha}
task, ensuring an appropriate sampling of the instrumental response.
In the 3--10~keV band, the source count rate ($\sim 1.8$~counts~s$^{-1}$) 
exceeds the estimated non-X-ray background (NXB) rate ($4.8 \times 10^{-3}$~counts~s$^{-1}$) 
by a factor of $\sim 370$ (i.e., the NXB contributes $\lesssim 0.3\%$ of the total).
Therefore, following \citet{Kayanoki+26}, the Resolve background spectrum was not subtracted.
The light curve of Resolve in the 2--10~keV band is plotted in the upper panel of Figure~\ref{fig:lightcurve}.
The Xtend data reduction was performed using the \texttt{xapipeline} task.
The Xtend data were taken with CCD1 and CCD2 in the 1/8 window mode.
We extracted the Xtend source spectrum from a rectangular region of 126 arcseconds $\times$ 360 arcseconds centered on the source position.
The source extraction region is contained within a single CCD chip.
The Xtend background spectrum was extracted from a source-free rectangular region of 90 arcseconds $\times$ 432 arcseconds on the other chip.
The redistribution matrix file was generated with the \texttt{xtdrmf} task.
The exposure map was generated with the \texttt{xaexpmap} task.
The ancillary response file for a point source was generated with the \texttt{xaarfgen} task.
The number of photons per raytraced energy in \texttt{xaarfgen} (\texttt{numphoton}) was set to $3\times10^{5}$.
The Xtend spectrum was binned to a minimum of 50 counts per bin using the \texttt{grppha} tool.
The effective exposure time of Xtend is 183~ks.
The light curve of Xtend in the 0.5--10~keV band is plotted in the middle panel of Figure~\ref{fig:lightcurve}.

\subsection{NuSTAR}\label{subsec:nustar}

\begin{figure}
  \centering
  \includegraphics[width=0.6\linewidth]{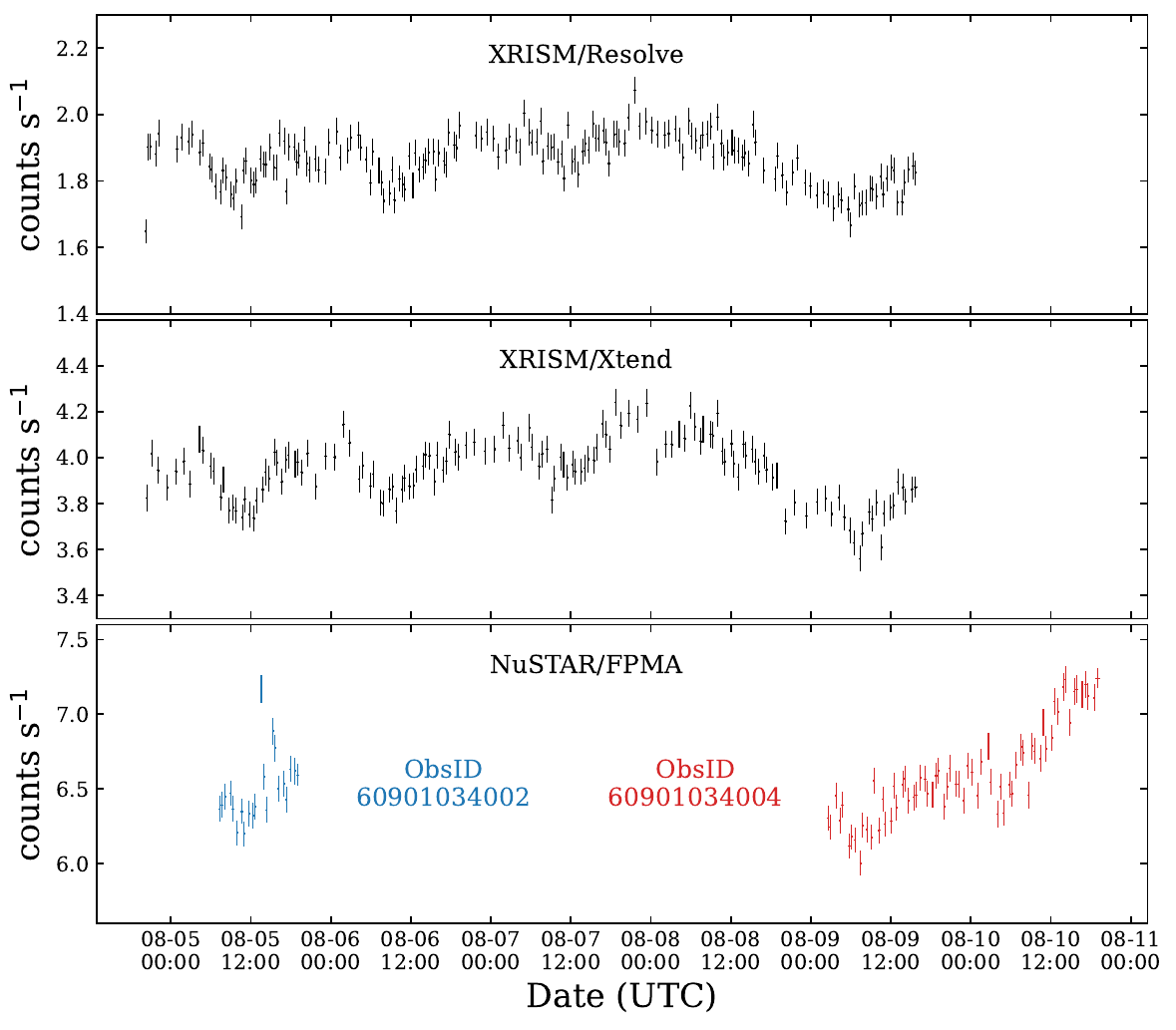}
  \caption{(Upper panel) \textit{XRISM}/Resolve light curve in the 2--10~keV band.
    (Middle panel) \textit{XRISM}/Xtend light curve in the 0.5--10~keV band.
    (Lower panel) \textit{NuSTAR}/FPMA light curves in the 3--78~keV band for the two observations, 
    ObsID 60901034002 (blue) and ObsID 60901034004 (red).
    The time bin is 1200~s for \textit{XRISM}/Resolve and \textit{NuSTAR}, and 1280~s for \textit{XRISM}/Xtend.
    The light curves for the FPMB detector are consistent with those of FPMA but are omitted for clarity.
  }
  \label{fig:lightcurve}
\end{figure}

To constrain the high-energy X-ray continuum and reflection
components, we also utilized a set of two \textit{NuSTAR} observations
performed simultaneously with the \textit{XRISM} observation (25 and 77~ks, respectively; Table~\ref{tab:obslog}). 
The data were processed using HEASoft version 6.37.1 with the
CalDB version 20260923.  We extracted the
source spectra from circular regions with radii of 120 arcseconds
centered on the source position for both the FPMA and FPMB detectors.
The background spectra were extracted from source-free circular
regions with radii of 120 arcseconds on the same detectors. We
generated the source and background spectra, along with the
corresponding response files, using the \texttt{nuproducts} script.
In order to improve the signal-to-noise ratio, the spectra from the two
observations were combined separately for FPMA and FPMB using the \texttt{addascaspec} tool.
The combined \textit{NuSTAR} spectrum was binned to have a minimum of 600 counts per bin using the \texttt{grppha} tool.
We first extended the \textit{NuSTAR} energy range up to 79~keV.
However, the bin containing 79~keV does not have enough statistics (a minimum of 600 counts per bin), and thus it is not used.
As a result, we use the \textit{NuSTAR} data up to $\sim$75~keV, where the source is still detected above the background.
Conservatively, we do not use the \textit{NuSTAR} data below 8~keV, 
given the possible cross-calibration uncertainties in the energy-dependent effective area relative to \textit{XRISM}/Resolve at these energies.
The light curve of \textit{NuSTAR}/FPMA in the 3--78~keV band for each observation is plotted 
in the lower panel of Figure~\ref{fig:lightcurve}, illustrating the overlap with the \textit{XRISM} observing window.

\section{X-ray Spectral Analysis}\label{sec:xrayanalysis}

In this section, we present our spectral analysis
of the simultaneously observed data of \textit{XRISM}/Resolve, \textit{XRISM}/Xtend, and \textit{NuSTAR}.
To model the spectra, we consider two distinct structures in the circumnuclear
region: a clumpy torus and a uniform disk-like xBLR in the equatorial plane, where the inner xBLR extends up to the torus.
First, we describe the simulations using the SKIRT code to create a torus reflection model tailored for Cen~A (Section~\ref{subsubsec:torus}), 
and a general xBLR reflection model (Section~\ref{subsubsec:blr}).
Next, we explain the overall spectral model used in our analysis
(Section~\ref{subsec:spectralmodel}).  Finally, we present the fitting
results with a primary focus on the modeling of the Fe K$\alpha$ line
profile (Section~\ref{subsubsec:fekalineprofile}).
All fits were performed by minimizing the Cash statistic (C-stat: \citealt{Cash+79}) using the XSPEC package (\citealt{Arnaud_1996}) version 13.0.1.
We simultaneously fitted the 
\textit{XRISM}/Resolve spectrum in the 3--10~keV band,
\textit{XRISM}/Xtend spectrum in the 3--6.0~keV and 7.7--10~keV bands,
and the \textit{NuSTAR}/FPMA and \textit{NuSTAR}/FPMB spectra in the 8--75~keV band.
For \textit{NuSTAR} data, we set the upper limit of the energy range to 75~keV because the bin containing 79~keV does not have enough statistics.
Here, we conservatively exclude the 6.0--7.7~keV band of the Xtend spectrum, 
which contains the Fe K$\alpha$, Fe K$\beta$, and Ni K$\alpha$ lines.
The line profiles in this band are determined by the Resolve data.
The Xtend data are used to constrain the continuum shape.
To account for possible cross-calibration uncertainties among the instruments, 
we introduce multiplicative constant factors for the \textit{XRISM}/Xtend and \textit{NuSTAR}/FPMA and FPMB spectra relative to the \textit{XRISM}/Resolve spectrum, 
as detailed in Section~\ref{subsec:spectralmodel}.

\begin{figure*}
  \centering
  \includegraphics[width=0.8\linewidth]{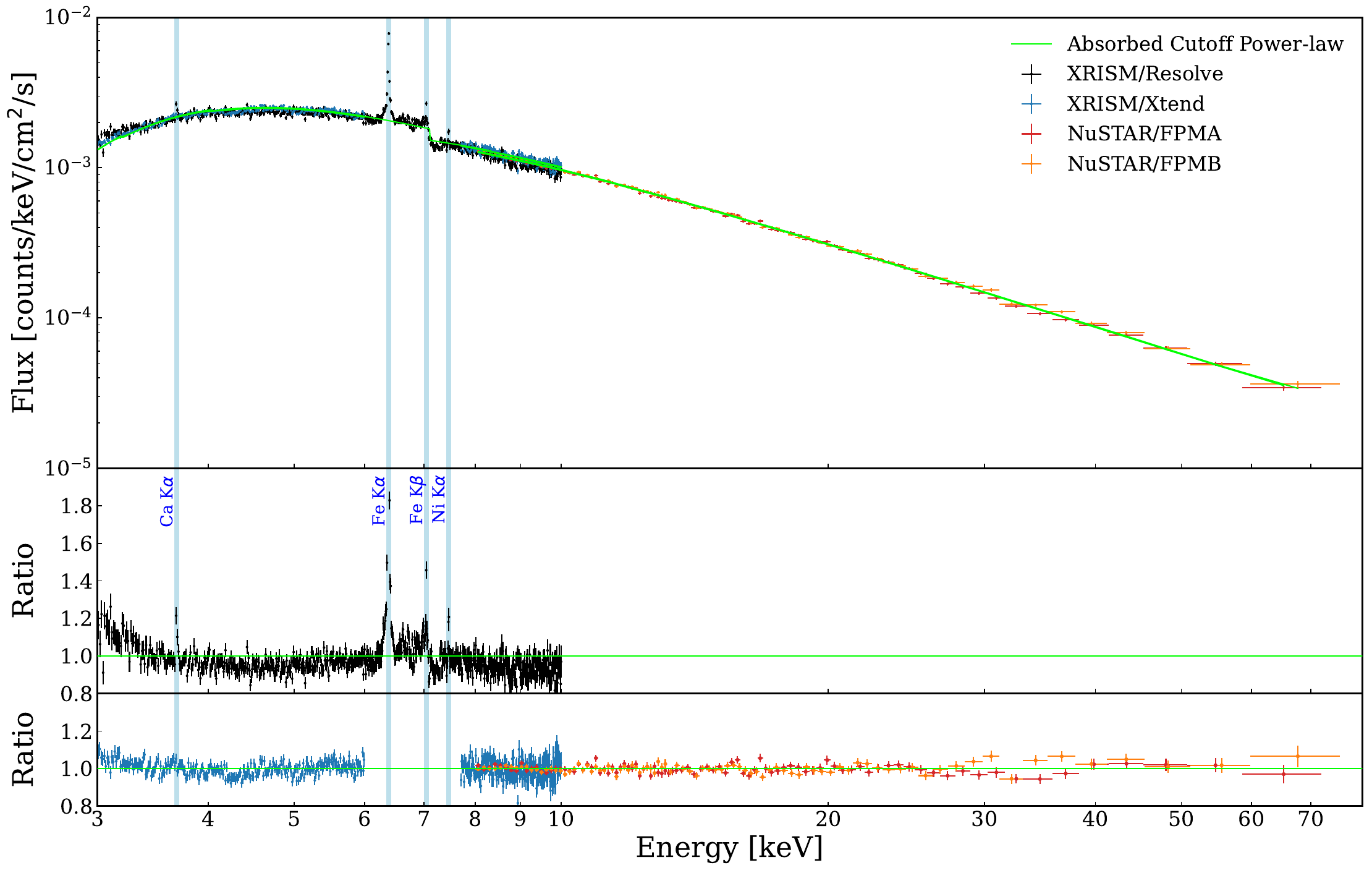}
  \caption{
    (Upper panel) Observed X-ray spectra folded with the energy responses (but corrected for effective area).
    Black, blue, red, and orange crosses represent the \textit{XRISM}/Resolve, \textit{XRISM}/Xtend, \textit{NuSTAR}/FPMA, and \textit{NuSTAR}/FPMB spectra, respectively.
    The model spectrum is shown as a green solid line.
    (Middle panel) Ratio of the Resolve spectrum to a simple absorbed cutoff power-law model.
    (Lower panel) Ratio of the Xtend and \textit{NuSTAR} spectra to the same model.
    The model is fitted simultaneously to all the spectra.
    The Xtend data in the 6--7.7~keV band are not used for the fit.
    The data points are rebinned for display.
    In all panels, the vertical blue lines mark the redshifted ($z = 1.858 \times 10^{-3}$) energies of the
    Ca K$\alpha$, Fe K$\alpha$, Fe K$\beta$, and Ni K$\alpha$ fluorescence lines.
  }
  \label{fig:CHratio}
\end{figure*}

Before presenting the detailed spectral model, we first illustrate its basic features
by fitting the broadband spectrum with a simple absorbed cutoff power-law model.
Figure~\ref{fig:CHratio} shows the observed broadband spectrum and the ratio of the observed spectrum to this simple model.
Two features stand out: a prominent Fe K$\alpha$ emission line, and a weak Compton reflection hump around 20--30~keV.
The soft excess below $\sim 4$~keV is attributable to the too simplified modeling, which 
will be reduced in the final model presented below (Figure~\ref{fig:plots}); the slight difference
in the ratios between Resolve and Xtend is due to the cross calibration uncertainties
at the lowest energies (\citealt{XRISM_NGC3783_II}).
The strong Fe K$\alpha$ line and the weak Compton reflection hump, which have long been known for Cen~A, motivate the detailed reflection modeling described in the following subsections.

\subsection{Physical Modeling of Reflection Components}\label{subsec:modelreflection}

We developed two
radiative transfer based numerical models representing the reflection components
from the torus and the xBLR. Since \textit{XRISM}/Resolve
enables us to resolve the detailed line profiles of fluorescent
lines with unprecedented accuracy, 
it is necessary to calculate models including realistic line profiles
with sufficiently fine energy grids that are compatible with the observed data.
For physically self-consistent interpretation, it is critical to
include the reflection continuum as well.
To calculate the reflection spectra, we performed Monte Carlo
simulations using the 3D radiative transfer code SKIRT (\citealt{Baes+03, VanderMeulen+23}). 
In our simulations, we assumed that all material
was neutral (i.e., not ionized). The physical processes considered in
the simulations include photoelectric absorption, Compton scattering
by bound electrons, and fluorescent line emission.

The two models were computed independently to minimize computational cost.
This means that
we neglect any potential interactions between the torus and the xBLR,
such as corrections of the incident spectrum to the torus
due to the absorption by the xBLR
and multiple scattering between the torus and xBLR regions.
We have confirmed that these effects are not significant in our case
as the angular width of the xBLR is much smaller than that of the
torus (see Section~\ref{subsubsec:fekalineprofile}). Because both the torus and
xBLR models are scale-invariant (i.e., the results depend only on the
relative geometry of the reflector, not on its absolute scale), we can
set their sizes independently when we apply the models to the observed
spectra. In the radiative transfer simulations, any velocity structure is ignored for
simplicity. The Doppler broadening effects will be taken into account
by convolving the resulting spectral models with Gaussian kernels.

\subsubsection{Torus}\label{subsubsec:torus}
We adopt a clumpy torus geometry of the CLUMPY model (\citealt{Nenkova+08a, Nenkova+08b}).
We follow its formulation in the XCLUMPY model (\citealt[][Section~2]{Tanimoto+19}).
The clumpy structure is modeled by distributing spherical clumps randomly.
Each clump has the same fixed size ($R_{\mathrm{clump}} = 0.002$~pc) and the same hydrogen number density ($n_\mathrm{H}^\mathrm{clump}$).
The clumps are distributed according to a power-law with a
radial density profile index ($q$),
and a Gaussian distribution with 
an angular width ($\Delta_{\mathrm{torus}}$).
Therefore, in the spherical coordinate system centered at the SMBH, the number density function of clumps $d(r, \theta, \phi)$ is given by (\citealt[][Equation~(1)]{Tanimoto+19})
\begin{equation}
  d(r, \theta, \phi) = N \left(\frac{r}{r_\mathrm{in}}\right)^{-q} \exp\left(-\frac{(\theta - \pi/2)^2}{\Delta_{\mathrm{torus}}^2}\right),
\end{equation}
where $N$ is the normalization factor and $r_\mathrm{in}$ is the inner radius of the torus.
The normalization $N$ is set by the mean number of clumps crossed by a straight path running outward 
from $r_\mathrm{in}$ to $r_\mathrm{out}$ in the equatorial plane ($N_\mathrm{clump}^\mathrm{Equ}$), 
which is given by
\begin{equation}
  N_\mathrm{clump}^\mathrm{Equ} = \int_{r_\mathrm{in}}^{r_\mathrm{out}} d(r, \frac{\pi}{2}, 0) \pi R_{\mathrm{clump}}^2 \mathrm{d}r,
\end{equation}
where $r_\mathrm{out}$ is the outer radius of the torus.
Thus, $N$ can be expressed as (\citealt[][Equation~(2)]{Tanimoto+19})
\begin{equation}
  N = \frac{(1-q) N_\mathrm{clump}^\mathrm{Equ}}{\pi R_{\mathrm{clump}}^2 r_\mathrm{in}^{q} (r_\mathrm{out}^{1-q} - r_\mathrm{in}^{1-q})}.
\end{equation}
The hydrogen column density along the equatorial plane is related to $n^\mathrm{clump}_\mathrm{H}$ as (\citealt[][Equation~(5)]{Tanimoto+19})
\begin{equation}
  N_\mathrm{H}^\mathrm{Equ} = \frac{4}{3} R_{\mathrm{clump}} \times n^\mathrm{clump}_\mathrm{H} \times N_\mathrm{clump}^\mathrm{Equ}.
\end{equation}
The term $\frac{4}{3} R_{\mathrm{clump}} \times n^\mathrm{clump}_\mathrm{H}$
represents the average column density of a randomly located single clump along a line of sight through it.

We fix $N_\mathrm{clump}^\mathrm{Equ} = 10$, $r_\mathrm{in} = 0.05$~pc, $r_\mathrm{out} = 1.00$~pc, and $q = 0.5$, following the XCLUMPY model (\citealt{Tanimoto+19}).
These values correspond to the mean values obtained by \citet{Ichikawa+15}, who applied the CLUMPY model to the infrared spectral energy distributions of 21 nearby AGNs.
Thus, the free geometrical parameters are $\Delta_{\mathrm{torus}}$ and $N_\mathrm{H}^\mathrm{Equ}$ (or equivalently $n^\mathrm{clump}_\mathrm{H}$).
We treated the abundances of calcium (Ca), 
and nickel (Ni) as variable parameters in the table model.
The abundances of the other elements were all fixed at the solar
values (\citealt{Lodders+09}), because they could not be constrained by our data due to the
limitation of band coverage or insufficient photon statistics.

In our SKIRT simulations for the torus, we configured the spatial
domain as a cube with sides of 40~pc. The computational grid utilizes
an octree-based adaptive mesh, with subdivision levels of 7 to 15.
This setup ensures that the smallest cell size is 40~pc$/ 2^{15}
\approx 0.001$~pc, allowing for sufficient spatial resolution to
resolve each clump structure.
The spectral energy grid spans 1.5--500~keV and was constructed as follows: from 1.5~keV to
10.0~keV, the grid is linearly divided with a step size of 0.5~eV, and
from 10~keV to 500~keV, the grid is divided into 500 bins on a
logarithmic scale.  The energy resolution of the grid for the soft
X-ray band is set to be higher than that of \textit{XRISM}/Resolve to
accurately reproduce the line profiles (matching the 0.5~eV grid of the \textit{XRISM}/Resolve response matrix).
The incident spectrum energy
distribution is modeled with a power-law with an exponential cutoff:
$I(E) \propto E^{-\Gamma} \exp(-E/E_\mathrm{cut})$, where $\Gamma$ is
the photon index and $E_\mathrm{cut}$ is the cutoff energy.  For each
simulation, $10^7$ photon packets were launched in total across the
full energy range, with the energy of each packet sampled from this
incident spectrum.  The parameter grid
for the torus model is summarized in Table~\ref{tab:modelgrid} (note
that the model is scale-free and absolute distances do not enter the results).

\begin{table*}
  \caption{Summary of Parameters.\label{tab:modelgrid}
    \newline
    (1) Torus angular width.
    (2) Hydrogen column density along the equatorial plane.
    (3) Photon index.
    (4) Abundance of calcium relative to hydrogen.
    (5) Abundance of nickel relative to hydrogen.
    (6) BLR angular width.
    (7) Hydrogen column density in the radial direction.
    (8) Photon index.
    (9) Abundance of calcium relative to hydrogen.
    (10) Abundance of nickel relative to hydrogen.
    (11) Inclination angle.
    (12) Cutoff energy.
    (13) Abundance of iron relative to hydrogen.
    \newline
  }
  \begin{tabular}{llll}
    \hline
    Note & Parameter & Grid Value & Units\\
    \hline
    Torus (XCLUMPY)\\
    (1) & $\Delta_{\mathrm{torus}}$ & 30, 35, 40, 45, 50, 55, 60, 65, 70 & deg \\
    (2) & $\log{N_\mathrm{H}^\mathrm{Equ}/\mathrm{cm}^{-2}}$ & 23.20, 23.30, 23.40, 23.50, 23.60, 23.70, 23.80, 23.90 & \nodata\\
    (3) & $\Gamma$ & 1.750, 1.775, 1.800, 1.825, 1.850 & \nodata\\
    (4) & $A_{\rm Ca}$ & 0.70, 0.90, 1.10, 1.30, 1.50, 1.70, 1.90 & \nodata\\
    (5) & $A_{\rm Ni}$ & 0.90, 1.00, 1.05, 1.20, 1.35, 1.50, 1.65 & \nodata\\
    \hline
    BLR (XBLR)\\
    (6) & $\Delta_{\rm BLR}$ & 6, 12, 18, 24, 30, 37, 45 & deg \\
    (7) & $N_{\rm H}^{\rm BLR}$ & 1, 5, 10, 30, 50, 80, 100, 130, 150, 200, 500, 1000 & $10^{22}$~cm$^{-2}$ \\
    (8) & $\Gamma$ & 1.750, 1.800, 1.850 & \nodata\\
    (9) & $A_{\rm Ca}$ & 0.70, 1.10, 1.90 & \nodata\\
    (10) & $A_{\rm Ni}$ & 0.90, 1.20, 1.65 & \nodata\\
    \hline
    Common Parameters \\
    (11) & $i$ & 50 & deg \\
    (12) & $E_\mathrm{cut}$ & 370 & keV \\
    (13) & $A_{\rm Fe}$ & 1.0 & \nodata\\
  \end{tabular}
\end{table*}

\subsubsection{xBLR}\label{subsubsec:blr}

For the xBLR model, we assume a partially spherical geometry with two
polar cuts between inner and outer radii, as adopted by
\citet{Stalevski+12} and \citet{Granato_Danese_1994}. We consider a small
angular width (i.e., small covering factor), assuming that the mass
distribution of the xBLR is concentrated close to the equatorial plane (\citealt{Czerny_Hryniewicz_2011}). 
The density distribution is expressed in spherical coordinates $(r, \theta)$ centered on the SMBH.
Here, $\theta$ is the polar angle measured from the symmetry axis of the xBLR, and the density is axisymmetric.
The equatorial plane corresponds to $\theta = 90^\circ$.
The density is given by
\begin{equation}\label{eq:xblrdensity}
  \rho(r, \theta) = A r^{-p_{\rm BLR}} e^{-q_{\rm BLR}|\cos{\theta}|} 
  \mathrm{~for~} r_{\mathrm{min}} \leq r \leq r_{\mathrm{max}} 
  \mathrm{~and~} |\theta - 90^\circ| \leq \Delta_{\mathrm{BLR}},
\end{equation}
where $A$ is the normalization factor,
$p_{\rm BLR}$ is the radial density profile index,
$q_{\rm BLR}$ is the angular density profile index,
$r_\mathrm{min}$ and $r_\mathrm{max}$ are the inner and outer radii of the xBLR, respectively,
and $\Delta_\mathrm{BLR}$ is the xBLR angular width (i.e., the half-opening angle of the xBLR from the equatorial plane).
We normalize the density using the hydrogen column density in the radial direction, $N_{\rm H}^{\rm BLR}$.
This is the column density along the equatorial plane, measured on one side of the SMBH from $r_{\rm min}$ to $r_{\rm max}$.
We set $r_\mathrm{min}=0.01$~pc, and assume the outer radius to be ten times the inner radius ($r_\mathrm{max} = 10\times r_\mathrm{min} = 0.1$~pc). 
For simplicity, we assume a uniform density within the xBLR by setting $p_{\rm BLR}=0$ and $q_{\rm BLR}=0$. 
Then, Equation~(\ref{eq:xblrdensity}) reduces to $\rho = A$ within the xBLR.
We regard $\rho$ as the hydrogen number density $n_{\rm H}$, and hence $n_{\rm H} = A$.
Because $n_{\rm H}$ is constant, $N_{\rm H}^{\rm BLR}$ is given by
\begin{equation}
  N_{\rm H}^{\rm BLR} = \int_{r_{\rm min}}^{r_{\rm max}} n_{\rm H}\, dr = n_{\rm H}\,(r_{\rm max} - r_{\rm min}).
\end{equation}
The normalization factor is therefore given by
\begin{equation}
  A = n_{\rm H} = \frac{N_{\rm H}^{\rm BLR}}{r_{\rm max} - r_{\rm min}} = \frac{N_{\rm H}^{\rm BLR}}{9\,r_{\rm min}}.
\end{equation}
The effects of clumpiness will be evaluated in Appendix~\ref{sec:appendix}.
The xBLR model has two free geometrical parameters:
the hydrogen column density in the radial direction ($N_{\rm H}^{\rm BLR}$), and the xBLR angular width ($\Delta_{\rm BLR}$).
The solid angle of the region $|\theta - 90^\circ| \leq \Delta_{\rm BLR}$ is $4\pi \sin\Delta_{\rm BLR}$.
The covering factor, which is the fraction of the sky covered by the xBLR as seen from the central SMBH, is therefore $\sin\Delta_{\rm BLR}$.
We note that this xBLR model is also a scale-free model.
As in the torus model,
the abundances of Ca, and Ni relative to H are set as
variable parameters.

In our SKIRT simulations for the xBLR, we configured the spatial domain as a cylinder with a radius of 0.11~pc and a height of 0.22~pc,
providing a small margin outside the physical outer radius ($r_{\mathrm{max}} = 0.1$~pc) to avoid computational boundary artifacts.
The spatial grid is a 2D cylindrical grid with 1000 radial bins and 1000 vertical bins.
This setup ensures that the edge of the xBLR is well resolved.
The number of photon packets launched in each simulation is $10^8$.
We used the same approach as with the torus model (Section~\ref{subsubsec:torus}) for the gas composition, the spectral energy grid, and the incident spectrum energy distribution.
The parameter grid for the xBLR model is summarized in Table~\ref{tab:modelgrid}.

\subsection{Spectral Model}\label{subsec:spectralmodel}

Our model represents a typical X-ray spectrum of an obscured AGN,
including the contribution from the xBLR, which has been mostly ignored
in previous X-ray studies. 
It consists of a primary power-law component with an exponential
cutoff from the central hot corona, its reflection component from the
xBLR, the emission component from photoionized plasmas, and the
reflection component from the torus. The first three components are
subject to the absorption by the dusty torus along the line of sight.

We find that the most prominent
observed spectral feature, Fe K$\alpha$ line, is modelled with
three components (see Section~\ref{subsubsec:fekalineprofile}): (1) the reflection from the xBLR that produces a
broad line, (2) and (3) two narrow line components from the torus with different velocity widths (torus2 and torus1, respectively)
To account for the kinematic broadening caused by the macroscopic motion of the gas (e.g., Keplerian rotation),
 we adopted a Gaussian smoothing kernel
(\textsf{gsmooth}) to broaden the components, instead of applying
a disk-line convolution kernel (such as \textsf{rdblur}).  The use of
a disk line kernel introduces degeneracy among the geometric
parameters; specifically, the inner radius ($r_{\text{in}}$) is
strongly coupled with the emissivity index $q$ and the outer
radius ($r_{\text{out}}$).
Fixing either or both of these parameters would impose artificial
biases in estimating $r_{\text{in}}$, rendering its physical meaning ambiguous.

In the XSPEC terminology, our model is described as follows:
\begin{eqnarray}
  \label{equation:model}
  \mathrm{Model} &=& \textsf{constant1}*\textsf{phabs} \nonumber\\
  &*& (\textsf{zvphabs}*\textsf{cabs} * (\textsf{zcutoffpl} + 3 * \textsf{atable\{PION\_CenA.fits\}} + \textsf{gsmooth1} * \textsf{atable\{xBLR\_CenA\_Resolve.fits\}}) \nonumber \\
  &+& \textsf{atable\{xclumpy\_CenA\_Resolve\_Continuum.fits\}} \nonumber\\
  &+& \textsf{constant2}*\textsf{gsmooth2}*\textsf{atable\{xclumpy\_CenA\_Resolve\_Lines.fits\}} \nonumber\\
  &+& (1 - \textsf{constant2})*\textsf{gsmooth3}*\textsf{atable\{xclumpy\_CenA\_Resolve\_Lines.fits\}}) \nonumber
\end{eqnarray}

\begin{enumerate}
  \renewcommand{\labelenumi}{(\arabic{enumi})}

\item The \textsf{constant1} term represents an instrumental cross-calibration constant among different detectors. 
  We set \textsf{constant1} of \textit{XRISM}/Resolve to unity as a reference.
  We allow those of \textit{XRISM}/Xtend, \textit{NuSTAR}/FPMA, and \textit{NuSTAR}/FPMB to vary independently.
  The \textsf{phabs} term represents the Galactic absorption.
  We fixed its hydrogen column density at $N^{\mathrm{Gal}}_{\mathrm{H}} = 1.17 \times10^{21}$~cm$^{-2}$, the total value of $HI$ and $H_{2}$ based on
  \url{https://www.swift.ac.uk/analysis/nhtot/} \citep{Willingale+13}.

\item The \textsf{zvphabs*cabs} term represents the photoelectric absorption and the extinction due to Compton scattering by the line-of-sight material. 
  Its hydrogen column density is calculated via Equation~(\ref{N_HLOS}) (see item (6) below) assuming that it originates only from the clumpy torus.
  As described in Section~\ref{subsubsec:blr}, the xBLR is assumed to be concentrated near the equatorial plane
  with an angular width smaller than $90^{\circ}-i$, and therefore does not intercept the line of sight.

\item The \textsf{zcutoffpl} term represents the primary power-law component from the hot corona. 
  Since our fits do not find the cutoff energy to be below 79~keV, and due to a lack of data above this energy, we fixed it to a typical value of 370~keV ($E_{\mathrm{cut}} = 370$~keV: \citealt{Ricci+18}).
  The free parameters of this component are the photon index ($\Gamma$) and the normalization ($K_{\rm PL}$).

\item The $3*\textsf{atable\{PION\_CenA.fits\}}$ terms represent the photoionized emission components.
  We adopted the same model as used in \citet{Kayanoki+26}, which consists of three components with different ionization parameters, 
  by converting their SPEX models into XSPEC table files.
  All parameters (hydrogen column density, ionization parameter, outflow velocity, and turbulent velocity) in each component and relative normalizations among the three components are fixed at the best-fit values reported in \citet{Kayanoki+26}.

\item The \textsf{gsmooth1} * \textsf{atable\{xBLR\_CenA\_Resolve.fits\}} terms represent the reflection component from the xBLR (Section~\ref{subsubsec:blr}). The \textsf{gsmooth1} term denotes the Doppler broadening effect.
  Its 1$\sigma$ width at 6.0~keV ($\sigma_{\rm broad}$) is set to be a free parameter, which is assumed to be proportional to the line center energy.
  The xBLR model parameters are: 
  the input power-law photon index,
  the hydrogen column density in the radial direction ($N_{\rm H}^{\rm BLR}$),
  the xBLR angular width ($\Delta_\mathrm{BLR}$),
  the abundances of
  calcium ($A_{\rm Ca}$),
  and nickel ($A_{\rm Ni}$),
  the redshift, and the normalization.
  We limited the xBLR angular width to $\Delta_{\rm BLR} \leq 40$~deg.
  For the assumed inclination of $i = 50$~deg, the line of sight is $90^{\circ} - i = 40$~deg above the equatorial plane.
  This limit therefore ensures that the line of sight does not cross the xBLR in our model.
  We note that, in this setting, $N_{\rm H}^{\rm BLR}$ is distinct from the line-of-sight column density ($N^{\rm LOS}_{\rm H}$) in item (2) above.

\item The XCLUMPY table, \textsf{atable\{xclumpy\_CenA\_Resolve\_Continuum.fits\}}, represents the reflection continuum without fluorescent lines from the torus (Section~\ref{subsubsec:torus}).
  The parameters of XCLUMPY are the input power-law photon index,
  the hydrogen column density along the equatorial plane
  $(N^{\rm Equ}_{\rm H})$, the torus angular width
  $(\Delta_{\mathrm{torus}})$, 
  the abundances of Ca ($A_{\rm Ca}$), 
  and Ni ($A_{\rm Ni}$), the redshift ($z$), and the normalization.
  The hydrogen column density along the line of sight ($N^{\rm LOS}_{\rm H}$)
  in item (2) is related to the torus parameters as follows:
  \begin{eqnarray} \label{N_HLOS}
    N^{\rm LOS}_{\rm H}\left(i\right) = N^{\rm Equ}_{\rm H} \exp\left(-\left(\frac{90^{\circ}-i}{\Delta_{\mathrm{torus}}}\right)^2\right).
  \end{eqnarray}
  In practice, the number of clumps intersecting the line of sight can fluctuate stochastically.
  Equation~(\ref{N_HLOS}) gives the expectation value of $N^{\rm LOS}_{\rm H}$, which we adopt in this work.

\item These two XCLUMPY tables, $\textsf{constant2}*\textsf{gsmooth2}*\textsf{atable\{xclumpy\_CenA\_Resolve\_Lines.fits\}}$ and
  $(1 - \textsf{constant2})*\textsf{gsmooth3}*\textsf{atable\{xclumpy\_CenA\_Resolve\_Lines.fits\}}$ represent the reflection fluorescent line components from the torus.
  As we will explain in Section~\ref{subsubsec:fekalineprofile},
  we found that the velocity profile of Fe K$\alpha$ is much better explained by two broadened components rather than a single one. 
  The \textsf{constant2} term, which we denote as $C_{\rm torus}$ ($0 < C_{\rm torus} < 1$), represents the fraction of the most narrow component (torus1) among the total torus component.
  The \textsf{gsmooth2} and \textsf{gsmooth3} terms represent the Doppler broadening effects for the two components.
  Their 1$\sigma$ widths ($\sigma_{\rm torus1}, \sigma_{\rm torus2}$) at 6.0~keV are allowed to vary with the physical constraint that $\sigma_{\rm torus2} > \sigma_{\rm torus1}$, both proportional to the line center energies.
  The other parameters of these two XCLUMPY line components are the same as those of the XCLUMPY continuum component described in (6) above.

\end{enumerate}

Here we summarize the details of our parameter settings. We fixed the inclination angle at 50~deg in all reflection
models (items (5), (6) and (7)) based on broadband X-ray studies using the
XCLUMPY model (\citealt{Ogawa+21}), in order to avoid degeneracy among torus geometrical
parameters. 
Varying $i$ within the range reported by \citet{Ogawa+21} does not significantly affect our results.
Since the data do not well constrain the abundance ratio
of iron with respect to hydrogen ($A_{\rm Fe}/A_{\rm H}$, see
Section~\ref{subsec:abundanceratio}), we fixed the Fe abundance at the
solar value. 
Fixing $A_{\rm Fe}/A_{\rm H}$ ties the absolute normalization of the reflection continuum to 
the fluorescent line fluxes at the solar iron abundance. 
Consequently, the Ca and Ni abundances relative to Fe are determined directly by 
the Ca K$\alpha$/Fe K$\alpha$ and Ni K$\alpha$/Fe K$\alpha$ line intensity ratios. 
These ratios are largely insensitive to the poorly constrained overall strength of the reflection continuum.
Since $A_{\rm Fe}$ is fixed at the solar value, 
the grid parameters $A_{\rm Ca}$ and $A_{\rm Ni}$ (defined relative to hydrogen in solar units; see Table~\ref{tab:modelgrid}) 
are numerically identical to the abundance ratios $A_{\rm Ca}/A_{\rm Fe}$ and $A_{\rm Ni}/A_{\rm Fe}$ discussed throughout the paper.
The redshift and the abundances of Ca and Ni are set to
be free but linked among the torus line-of-sight column density (item
(2)) and all reflection models (items (5), (6), and (7)). The incident
photon index, cutoff energy, and normalization in the reflection
models (items (5), (6), and (7)) are linked to those of the direct
component (\textsf{zcutoffpl}) in item (3).

\subsection{Fe K$\alpha$ Line Modeling}\label{subsubsec:fekalineprofile}

\begin{figure}
  \centering
  \includegraphics[width=0.8\linewidth]{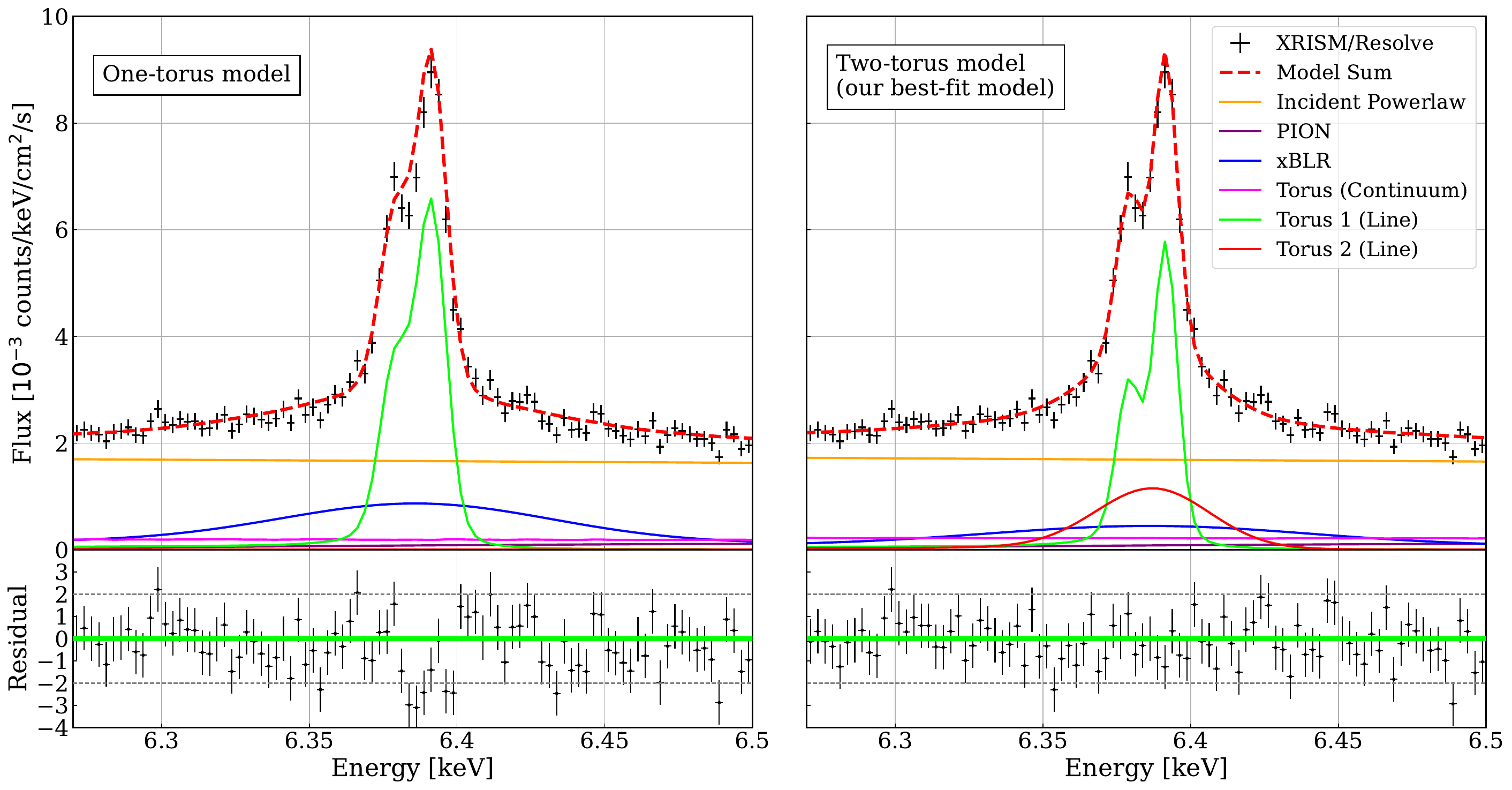}
  \caption{ (Left panel) Zoom-in on the Fe K$\alpha$ line with the ``one torus'' model ($\textsf{constant2} = 1.0$).
    Black crosses represent the \textit{XRISM}/Resolve observed spectrum folded with the energy responses (but corrected for effective area).
    The dashed red line represents the best-fit model, and the colored solid lines show the contributions of each model component:
    direct component (orange), photoionized component (purple), 
    reflection component from the xBLR (blue),
    reflection continuum from the torus (magenta), 
    reflection emission lines from torus1 (green).
    (Right panel) The same plot but for the ``two torus'' model ($\textsf{constant2}$ is free).
    The solid red line represents the contribution of the second torus (torus2) component, which has a broader velocity width than the torus1 component.
    The bottom sub-panels for both models show the fitting residuals in units of $\sigma$, plotted on the same vertical scale and with the same binning.
    The dashed gray lines indicate $\pm2\sigma$ levels.
  }
  \label{fig:FeZoom}
\end{figure}

Firstly, we fitted the spectrum with a single velocity width for the
torus component.  We fixed $\textsf{constant2} = 1.0$.  
In this modeling, we assumed that two reflection
components, the torus and the xBLR, have different velocity widths.
Their best-fit FWHMs were found to be ${\rm FWHM}_{\rm torus} =
430\pm40$~km~s$^{-1}$ and ${\rm FWHM}_{\rm BLR} =
5000\pm500$~km~s$^{-1}$, for the torus and the xBLR components,
respectively.  
These widths are consistent with those obtained by \citet{Bogensberger+25} from the same \textit{XRISM} data. 
They found a FWHM of $(480\pm20)$~km~s$^{-1}$ for the narrow cores and $(4300\pm300)$~km~s$^{-1}$ for the broad component. 
The fit statistic was $C$ = 
4179.04 for 3906 degrees of
freedom (d.o.f.).  However, the single velocity torus model failed to
reproduce the resolved shallow peak feature of Fe K$\alpha$ line,
leaving residuals in the 6.4~keV region, as shown in
Figure~\ref{fig:FeZoom}.  To account for this, we introduced a second
component for the torus reflection with a different velocity width.

Secondly, we set $\textsf{constant2}$ as a free parameter, allowing both 
$\textsf{constant2}*\textsf{gsmooth2}*\textsf{atable\{xclumpy\_CenA\_Resolve\_Lines.fits\}}$ and $(1 - \textsf{constant2})*\textsf{gsmooth3}*\textsf{atable\{xclumpy\_CenA\_Resolve\_Lines.fits\}}$ components 
to contribute to the torus reflection.
Here, we note that the total intensity from the torus reflection is still conserved.
This treatment of splitting the torus reflection into two velocity components is for testing whether a non-uniform torus provides a better description of the data.
With this modeling, we obtained the best-fit FWHMs of ${\rm FWHM}_{\rm torus1} = 
310\pm50$~km~s$^{-1}$, ${\rm FWHM}_{\rm torus2} = 
1900\pm400$~km~s$^{-1}$, and ${\rm FWHM}_{\rm BLR} = 
6000_{-800}^{+1100}$~km~s$^{-1}$, 
for the two torus components and the xBLR component, respectively.
The fit statistic improved from $C$ = 
4179.04 (3906 d.o.f.) to $C$ = 
4126.31 (3904 d.o.f.), corresponding to $\Delta C = 
52.73$ for two additional free parameters ($\Delta$d.o.f. $= 2$).
To evaluate the statistical significance of adding the second torus component, we calculated the Akaike Information Criterion (AIC; \citet{Akaike_1974}) and the Bayesian Information Criterion (BIC).
For the single-torus model, we obtained $\text{AIC} = 
4209.0$ and $\text{BIC} = 4303.1$.
With the two-torus model, these values improved to $\text{AIC} = 
4160.3$ and $\text{BIC} = 4267.0$. 
The substantial reductions of $\Delta \text{AIC} = 
-48.7$ and $\Delta \text{BIC} = -36.1$ support the necessity of the dual velocity components.
Furthermore, the systematic residuals seen around the Fe~K$\alpha$ line in the single-velocity model (Figure~\ref{fig:FeZoom}) are fully resolved by including this second torus component.
To check the goodness of fit, we binned the Resolve and Xtend spectra to have at least 300 counts per bin and performed the $\chi^2$ fitting.
The reduced chi-square value was $\chi^2/$d.o.f. $= 
2810.02/2504 \sim 1.12$, indicating a reasonable fit without significant systematic residual structures, given the high statistical quality of the data.

\begin{figure*}
  \centering
  \includegraphics[width=0.8\linewidth]{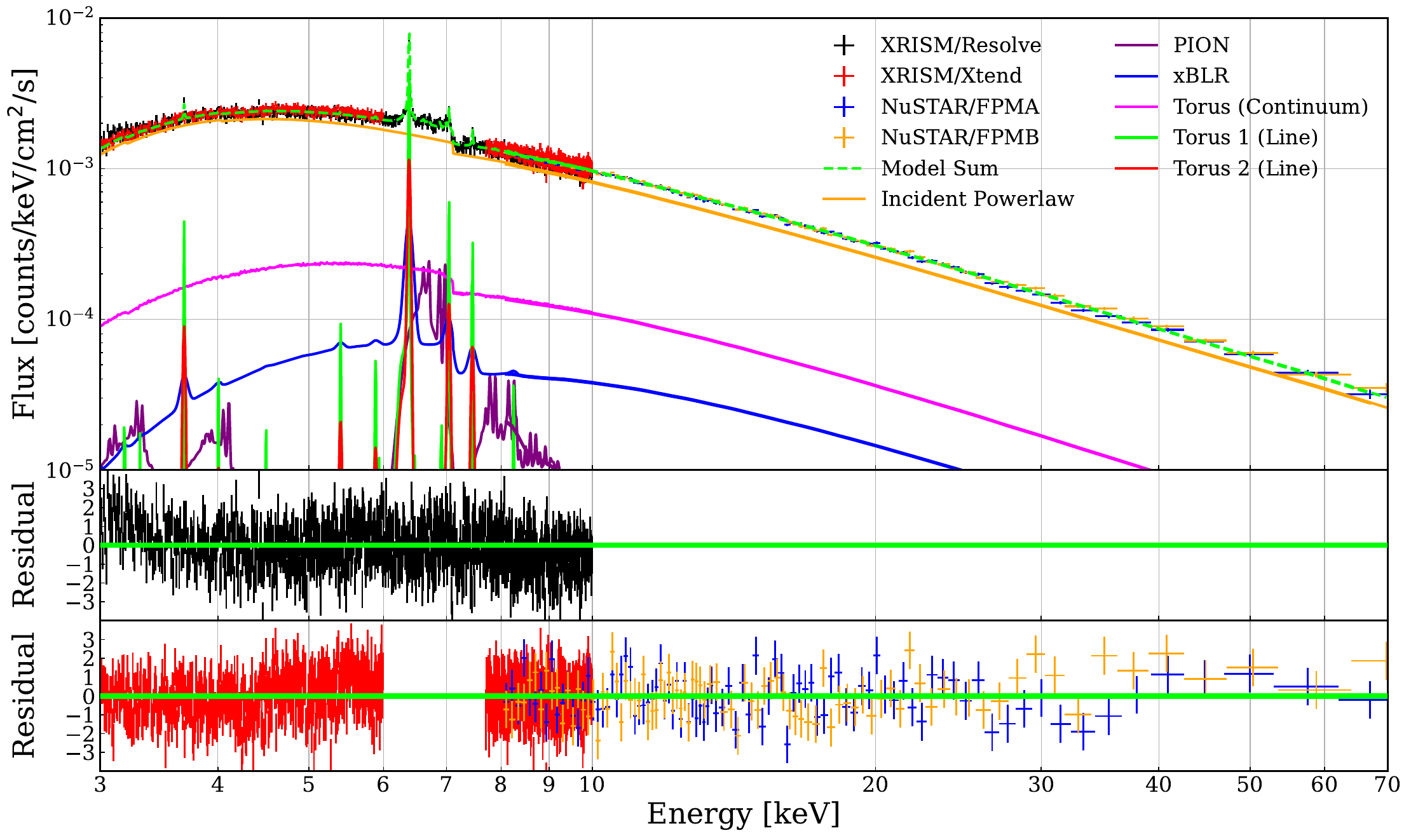}
  \caption{(Upper panel) Observed X-ray spectra folded with the energy responses (but corrected for effective area): 
    Black, red, blue, and orange crosses represent the
    \textit{XRISM}/Resolve, \textit{XRISM}/Xtend, \textit{NuSTAR}/FPMA, and \textit{NuSTAR}/FPMB spectra, respectively.
    The Xtend data in the 6--7.7~keV band are not used for the fit.
    The dashed lime line represents the best-fit ``two torus'' model for the broadband fitting.
    The other colored solid lines show the contributions of each model component, in the same way as in the right panel of Figure~\ref{fig:FeZoom}.
    The model curves are shown only for Resolve and \textit{NuSTAR}.
    (Middle panel) Fitting residuals of the \textit{XRISM}/Resolve spectrum in units of $1\sigma$ error.
    (Lower panel) Same as the middle panel, but for the \textit{XRISM}/Xtend and \textit{NuSTAR} (FPMA and FPMB) spectra.
    For display purposes, the data points in all panels are rebinned.
  }
  \label{fig:plots}
\end{figure*}

Thus, our final model includes three reflection components: two velocity components for the torus reflection component and one for the xBLR.
The best-fit parameters are summarized in Table~\ref{tab:best}, and the fit results are shown in Figure~\ref{fig:plots}.
\begin{table*}
  \caption{Best Fit Parameters.\label{tab:best}
    \newline
    (1) Abundance ratios of calcium and nickel to iron.
    (2) Photon index of the direct component and 
    the normalization in 10$^{-3}$~photons~keV$^{-1}$~cm$^{-2}$~s$^{-1}$ at 1~keV. 
    (3) Hydrogen column density of the xBLR in the radial direction in 10$^{22}$~cm$^{-2}$,
    the angular width of the xBLR in degree, and
    the velocity dispersion of the xBLR in eV.
    (4) Hydrogen column density along the equatorial plane in 10$^{22}$~cm$^{-2}$,
    torus angular width in deg,
    the fraction between the torus1 in the total torus component,
    hydrogen column density along the line of sight in 10$^{22}$~cm$^{-2}$,
    and the Doppler broadening sigma at 6.0~keV in eV for torus1 and torus2 components, respectively.
    (5) Intrinsic (absorption-corrected) photon flux of the \textsf{PION} component integrated over 6.0--8.0~keV, in units of 10$^{-4}$~photons~cm$^{-2}$~s$^{-1}$.
    (6) Redshift.
    (7) Instrumental cross-calibration constants, 
    relative normalization of \textit{XRISM}/Xtend to \textit{XRISM}/Resolve, \textit{NuSTAR}/FPMA to \textit{XRISM}/Resolve, and \textit{NuSTAR}/FPMB to \textit{XRISM}/Resolve.
    (8) C-stat and degrees of freedom (d.o.f.) values of the best-fit models.
  }
  \centering
  \begin{tabular}{ccclc}
    \hline 
    & Model &   & & Units\\
    \hline 
    (1) & Abundance & $A_{\mathrm{Ca}}/A_{\mathrm{Fe}}$& $1.10_{-0.11}^{+0.20}$ &  \\
    &  & $A_{\mathrm{Ni}}/A_{\mathrm{Fe}}$ &  $1.35_{-0.18}^{+0.07}$  &\\
    (2) & \textsf{zcutoffpl} & $\Gamma$  & $1.804_{-0.007}^{+0.010}$   & \\
    &  & $K_{\mathrm{PL}}$  & $69.3_{-1.4}^{+1.7}$  & see caption\\
    (3) & \textsf{xBLR} & $N_{\rm H}^{\rm BLR}$ & $50_{-31}^{+51}$ & 10$^{22}$~cm$^{-2}$\\
    &  & $\Delta_\mathrm{BLR}$ & $10.4_{-4.1}^{+11}$ & deg \\
    &  & $\sigma_{\rm broad}$ & $50.6_{-6.6}^{+8.9}$   & eV~at~6~keV \\
    (4) & \textsf{XCLUMPY} & $N^{\rm Equ}_{\rm H}$  & $33.4_{-3.0}^{+6.3}$ & 10$^{22}$~cm$^{-2}$\\
    &  & $\Delta_{\mathrm{torus}}$ & $44_{-4}^{+3}$  & deg\\
    &  & $C_{\rm torus}$ &  $0.64_{-0.08}^{+0.06}$  & \\
    &  & $N^{\rm LOS}_{\rm H}$  &  $14.57_{-0.12}^{+0.14}$ & 10$^{22}$~cm$^{-2}$ \\
    &  & $\sigma_{\rm torus1}$  &  $2.62_{-0.45}^{+0.39}$ & eV~at~6~keV\\
    &  & $\sigma_{\rm torus2}$  & $16.5_{-3.3}^{+3.5}$ & eV~at~6~keV \\
    (5) & \textsf{PION} & $K_{\mathrm{PION}}$ & $1.49_{-0.18}^{+0.17}$ & see caption \\
    \hline
    (6) & & $z$& $1.858_{-0.041}^{+0.039}$ &  10$^{-3}$ \\
    (7) & \textsf{constant} & $C_{\mathrm{Xtend}/\mathrm{Resolve}}$& $1.021_{-0.004}^{+0.004}$  &   \\
        & \textsf{constant} & $C_{\mathrm{FPMA}/\mathrm{Resolve}}$& $1.024_{-0.005}^{+0.005}$  &   \\
        & \textsf{constant} & $C_{\mathrm{FPMB}/\mathrm{Resolve}}$&  $1.037_{-0.005}^{+0.006}$ &   \\
    (8) &  & C-stat &  4126.31   & \\
    &  & d.o.f. & 3904  & \\
    \hline
  \end{tabular}
\end{table*}

\section{Discussion}\label{sec:discussion}

\subsection{Torus Structure of Cen A}\label{subsubsec:torusstructure}

We investigate the torus structure of Cen~A, on the basis of our
fitting results of the torus component modeled via XCLUMPY. Through
the broad-band analysis utilizing the simultaneous \textit{XRISM}/Resolve, \textit{XRISM}/Xtend, and
\textit{NuSTAR} data, we obtained the torus angular width of
$\Delta_{\mathrm{torus}} = 
44_{-4}^{+3}$~deg and the hydrogen column
density in the equatorial plane of $N_{\rm H}^{\rm Equ} =
33.4_{-3.0}^{+6.3}\times 10^{22}$~cm$^{-2}$ for an assumed inclination of 50
deg. These values are consistent with the previous results
obtained by \citet{Ogawa+21} based on the simultaneous analysis of a \textit{NuSTAR}
spectrum and \textit{Suzaku}/XIS ones.

Thus, the broadband continuum and high-resolution X-ray spectra point to a picture 
in which the nucleus in Cen~A is surrounded by a torus that is geometrically extended (having a large solid angle) but globally Compton-thin. 
The estimated covering fraction of the torus with $\log N_{\rm H}/{\rm cm}^{-2}>22$ is close to unity, 
in agreement with the relation between the Eddington ratio and torus covering fraction reported by \citet{Ricci+17Nature}. 
The absence of broad line components in either the direct or polarized optical spectra of Cen~A may be explained by this ``deeply buried'' nature of the nucleus.
Crucially, such a highly developed solid angle is also independently required to produce the prominent Fe~K$\alpha$ emission lines observed by \textit{XRISM}/Resolve, under the condition of the globally low column density of the reflecting matter.

This globally Compton-thin nature is strictly constrained by the \textit{NuSTAR} data. 
As shown in Figure~\ref{fig:CHratio}, there is no significant excess observed around 20--30~keV, 
and we obtain an upper limit of $\approx 5$\% on the hump strength relative to the continuum, confirming quantitatively the long-known weakness of the reflection in Cen~A.
Under our assumed model, the observed data are well explained by the relevant Thomson optical depths of $\tau_{\rm T} = \sigma_{\rm T}N_{\rm H} \approx 0.1$ along the line of sight, 
$0.2$ along the equatorial plane of the torus, and only $0.02$ through a single torus clump.

The necessity of evaluating the global gas distribution, rather than relying solely on the line-of-sight absorption, is well demonstrated by Mrk~3.
\citet{Yaqoob_2015} showed that self-consistent modeling of the Fe~K$\alpha$ line and the reflection continuum 
instead implies a global (equatorial) column density well below the Compton-thick threshold, 
illustrating that the line-of-sight absorption alone does not necessarily reflect the properties of the reprocessing matter as a whole.
The high energy resolution of \textit{XRISM}/Resolve provides further support for this insight. 
Furthermore, Mrk~3 operates at a low accretion rate ($\log\lambda_{\rm Edd}\sim-2.21$; \citealt{Koss+22}).
Given that Cen~A is also a low-Eddington-ratio AGN, such globally Compton-thin but geometrically extended tori could potentially represent a common feature among this population, raising the possibility of a wider diversity in torus properties than commonly assumed.

We note that whereas \citet{Bogensberger+25}
obtained somewhat lower inclinations of $24_{-7}^{+13}$ degrees by fitting
the Fe K$\alpha$ profile with the MYTORUS model (\citealt{Murphy_Yaqoob_2009}) broadened by an \textsf{rdblur} kernel, the profile
is still consistent with a high inclination up to 90 degrees within
the 90\% confidence limit. 
However, when incorporating constraints from reverberation mapping, high inclinations are ruled out (\citealt{Bogensberger+25}).
The low inclination is also consistent with the jet proper motion measurements 
($i < 47$ degrees as a $1\sigma$ upper limit; \citealt{Bogensberger+24}), 
where slightly higher angles (e.g., $\sim 50$ degrees) are only mildly disfavoured, 
and any minor discrepancy can be naturally accommodated by a disk-torus misalignment.

We have found that the profile of Fe K$\alpha$ component from the torus is
well characterized by two broadened components with different velocity
widths, $\sigma_{\rm torus1} = 
2.6\pm0.4$~eV and $\sigma_{\rm torus2}
= 16.5\pm3.5$~eV, which are converted to velocity dispersions in FWHM
of $v_{\rm FWHM} =
310\pm50$~km~s$^{-1}$ and $v_{\rm FWHM}= 
1900\pm400$~km~s$^{-1}$,
respectively. The line width reported in \citet{Bogensberger+25} based on
a single Gaussian fit, $\sim 500$ km s$^{-1}$, falls between these
values.
It is noteworthy that the two-velocity component models are also found to well
reproduce the Fe K$\alpha$ profiles observed with \textit{XRISM} in the Seyfert
galaxies NGC~4388 (\citealt{Fujiwara+26}) and NGC~4507 (Aiso et al. in prep.),
which have Eddington ratios of 0.1 (\citealt{Fujiwara+26}) and 0.07 (\citealt{Koss+22}), respectively. 
Thus, the two-velocity component models (i.e., three component model including the
xBLR component) may give a common description of the line profile in
obscured AGNs over a wide range of Eddington ratio.
Cen~A is a radio galaxy, while NGC~4388 and NGC~4507 are radio-quiet.
\citet{Nakatani+23} found that the geometric structure of the torus 
depends on the Eddington ratio, not on the presence of a jet.
The comparison above is therefore not affected by the different radio properties of these objects.
Following these works, we hereafter refer to the torus1 and torus2
components as ``main torus'' and ``inner torus'' regions,
respectively.

Assuming Keplerian motion, we estimate representative distances of the
main and inner torus regions from the SMBH to be $r \sim 
2\pm1 \times 10^{6}~r_{\rm g}\sim 5.3_{-1.3}^{+2.4}$~pc
and $r \sim 
6_{-2}^{+3} \times 10^{4}~r_{\rm g} \sim 0.13_{-0.04}^{+0.07}$~pc ($r_{\rm g} \equiv GM_{\rm BH}/c^2$ is the
gravitational radius where $G$ is the gravitational constant, $M_{\rm
  BH}$ is the black-hole mass, and $c$ is the light speed) from the
observed line widths, respectively. Here we used the relation
\begin{equation}\label{eq:fwhm2rg}
  \frac{r}{r_{\rm g}} = \left(\frac{c}{v_{\rm FWHM}/(2\sin i)}\right)^2,
\end{equation}
where $i$ is the inclination angle of the system.
We adopted $i = 50$~deg, which is the value fixed in our reflection models (Section~\ref{subsec:spectralmodel}).
The estimated distances scale as $\sin^2 i$.
These distances are
consistent with those of two iron-line emitting regions suggested by
\citet{Iwata+24} based on the reverberation mapping of the Fe K$\alpha$ line
using the \textit{NuSTAR}, \textit{Suzaku}, \textit{XMM-Newton}, and \textit{Swift}/XRT data with the
\textit{Swift}/BAT light curve.
We note that, in general, the two-velocity component model is a
phenomenological description of the data and does not necessarily
indicate that the emitter actually consists of two ``distinct''
regions. If a single smooth region is located over a wide range of
radius with comparable intensities, the observed profile should be
better represented by multiple velocity components. It is very
difficult to distinguish which (distinct or smooth matter distribution)
is the case, solely based on the line profile. In fact, we find that
the narrow line profile can be also reasonably fit with a \textsf{rdblur}
kernel,
yielding the inner and outer radii of $r_{\rm in} = 
2\pm1 \times 10^{4}~r_{\rm g}$ and $r_{\rm out} = 4_{-2}^{+3}\times 10^{6}~r_{\rm g}$
for a fixed emissivity index of $q=-2$.

A notable difference in torus structure between Cen~A and 
the higher-Eddington-ratio Seyfert~2s NGC~4388 and NGC~4507 
lies in the relation between the dust sublimation radius ($r_{\rm sub}$) and the torus inner radius. 
In Cen~A, $r_{\rm sub}$ is estimated to be 0.013~pc from equations~(10) and~(11) of \citet{Kudoh+24}, 
one order of magnitude smaller than the torus inner radius. 
In NGC~4388 and NGC~4507, by contrast, $r_{\rm sub}$ and the torus inner radius are comparable to each other. 
This gap between $r_{\rm sub}$ and the torus inner radius in Cen~A implies that dust can exist in its inner region, 
but the inner edge of the dense torus itself is not set by dust sublimation physics, 
such as radiation pressure on dust (e.g., \citealt{Wada+16}). 
Instead, in such low-luminosity (low Eddington-ratio) systems, 
this torus inner edge is likely determined by other (unknown) large-scale processes, 
such as the total gas mass in the circumnuclear region. 
Further studies using a large sample covering a wide range of Eddington ratio are crucial 
to reach a unified view on the structure of AGN tori and their evolutionary scenario.

Recent \textit{JWST} spectroscopic observations of the Cen~A nucleus have provided new constraints on the line profiles 
in the near-infrared (NIR) band.
Given that the torus is composed of molecular gas and dust,
the Fe K$\alpha$ reflection in X-rays and the molecular emission lines in the NIR should trace the kinematics of the same material.
\citet{Goold+26} reported multiple vibrational-rotational transitions of H$_2$ emission lines. 
The (1-0)O(5) line at 3.235~$\mu$m and the (1-0)O(6) line at 3.501~$\mu$m have 
FWHM$=2250\pm720$~km~s$^{-1}$ and $160\pm90$~km~s$^{-1}$, respectively (both detected at $S/N\sim4$).
These measurements agree, within the uncertainties, with the velocity widths of the inner torus component
(torus2, FWHM$=
1900\pm400$~km~s$^{-1}$) and the main torus component (torus1, FWHM$=
310\pm50$~km~s$^{-1}$) 
obtained in this study, respectively.
We note that H$_2$ emission is strongly affected by dust extinction, 
and thus it is not obvious whether the inner torus region can be probed as in the X-ray band.
To estimate the extinction, we use the line-of-sight column density of 
$N_{\rm H}^{\rm LOS}=
14.6\pm0.2\times10^{22}$~cm$^{-2}$ obtained in this X-ray observation.
Assuming the standard Galactic gas-to-dust ratio, this corresponds to $A\sim5$~mag at 3.235~$\mu$m,
i.e., a flux suppression by a factor of $\sim100$.
The observed H$_2$ line fluxes may therefore represent only a fraction of the total emission from the torus,
and deeper regions could remain unprobed by these NIR lines.
We note, however, that the signal-to-noise ratios of the H$_2$ detections are low (3.8--4.0), and further confirmation with higher-sensitivity infrared spectroscopy is needed to establish this correspondence more robustly.

\subsection{Broad Line Component of Fe K$\alpha$: X-ray BLR}\label{subsec:broadfeka}

In this subsection, we discuss the structure and physical origin of
the broad line component of Fe K$\alpha$ in Cen A on the basis of our
spectral analysis results employing the xBLR reflection model (see
Table~\ref{tab:best} for the best-fit parameters). The velocity width
of the xBLR component was found to be $\sigma_{\rm BLR} =
50.6_{-6.6}^{+8.9}$~eV, which is converted to ${\rm FWHM}_{\rm BLR} =
6000_{-800}^{+1100}$~km~s$^{-1}$. We obtained $\Delta_\mathrm{BLR} = 
10.4_{-4.1}^{+11}$ deg and $N_{\rm H}^{\rm BLR} = 
50_{-31}^{+51} \times 10^{22}$~cm$^{-2}$.
In addition to characterizing the xBLR itself, 
we also demonstrate how the precise measurement of its velocity width 
can be applied to estimate the SMBH mass in obscured AGNs more generally.

\subsubsection{Physical Structure of the X-ray BLR}\label{subsubsec:physicalstructurexrayblr}

\begin{figure}
\centering
\includegraphics[width=0.6\linewidth]{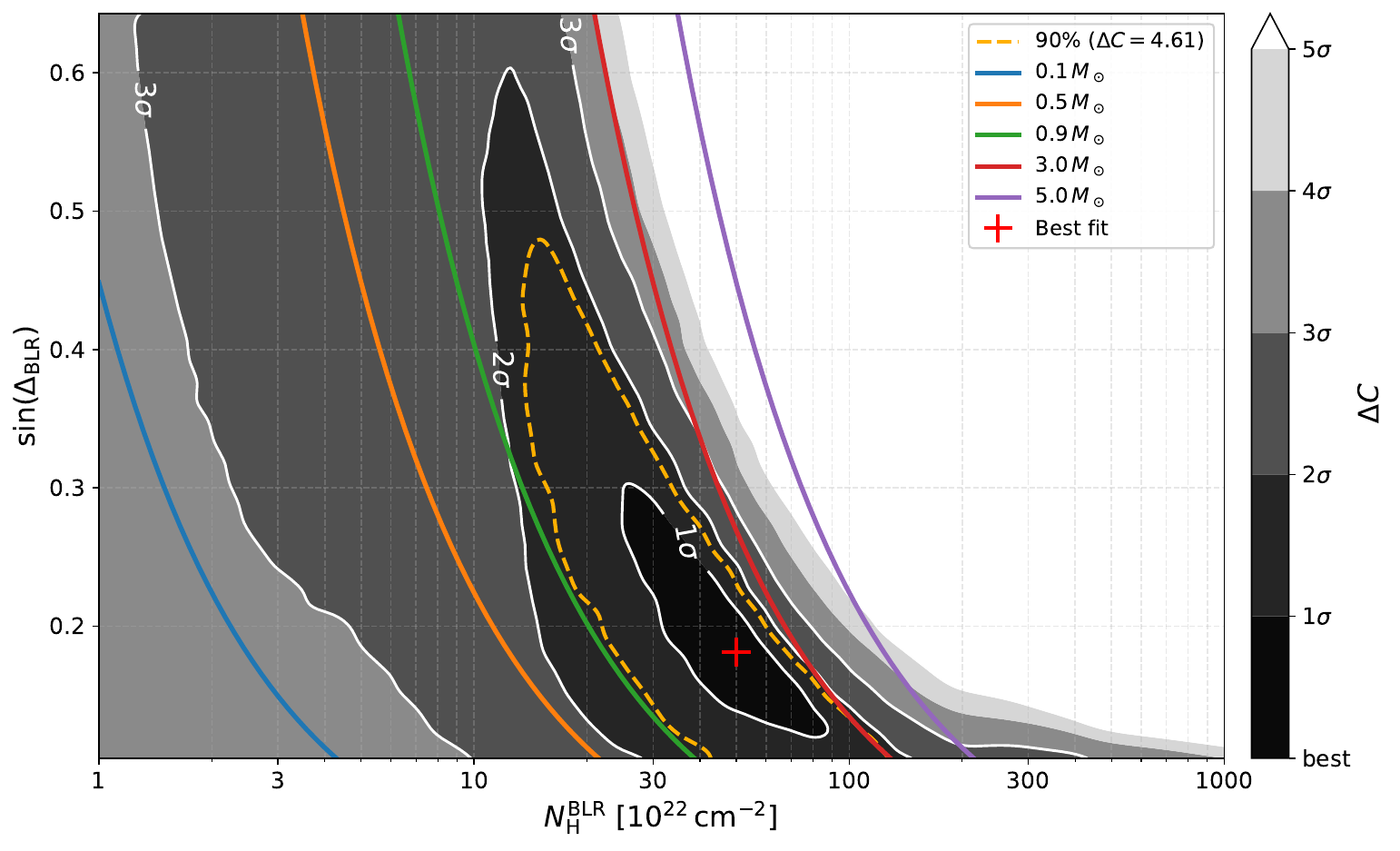}
\caption{
  Confidence contours of the xBLR angular width and radial column density.
  The color map shows $\Delta C$, the $C$-statistic relative to its best-fit value.
  The white curves show the 1, 2, and 3$\sigma$ contours.
  The yellow dashed curve shows the joint 90\% confidence region for two parameters ($\Delta C = 4.61$).
  The red cross marks the best-fit position.
  The solid blue, orange, green, red, and purple curves represent constant xBLR mass (0.1, 0.5, 0.9, 3.0, and 5.0~$M_{\odot}$, respectively).
  These curves are calculated for the shell geometry described in Section~\ref{subsubsec:massblr}, with $r_{\rm BLR}$ and $\Delta r_{\rm BLR}$ fixed at their best-fit and derived values.
}
\label{fig:HOA-NHContour}
\end{figure}

The xBLR angular width (
6.3--21 degrees) corresponds to a covering
factor of $ 
0.11 < C_{\rm f} < 0.36$ (see
Section~\ref{subsubsec:blr}). Similarly to the case of the torus
reflection, the lack of a strong reflection hump feature around
20$\textendash$30~keV suggests that the xBLR also has a low column
density.  Figure~\ref{fig:HOA-NHContour} shows the confidence contours
of the xBLR angular width and the radial column density. As noticed,
to reproduce the observed intensity of the broad Fe K$\alpha$ line
without a strong reflection hump, the xBLR must have a large covering
factor and a low column density.

Assuming that the xBLR gas is in Keplerian rotation around the SMBH
with the kinematic mass of $M_{\rm BH} = 10^{7.7}~M_{\odot}$ (Section~\ref{sec:intro}), we estimate the distance of
the xBLR from the SMBH to be $r_{\rm BLR}(\rm{Fe~K}\alpha) =
5900_{-1700}^{+1900}~r_{\rm g} \sim 1.4\pm0.4 \times 10^{-2}$~pc, according to
Equation~(\ref{eq:fwhm2rg}).
This value is close to the dust sublimation radius $r_{\rm sub} = 1.3\times 10^{-2}$~pc,
implying that dust-related physics (such as dust opacity) is relevant to produce this structure. 
Note that dust can survive at the surface of the
accretion disk even at smaller radii than $r_{\rm sub}$, and hence common mechanisms may also work in higher luminosity AGNs than Cen A.

The center energy of the Fe~K$\alpha$ line is consistent with 6.4~keV, 
indicating that the xBLR is mainly composed of neutral material. 
The ionization parameter is defined as $\xi \equiv L_{\rm X}/nr^2$ (\citealt{Tarter+69}), 
where $L_{\rm X}$ is the luminosity, $n$ the density, and $r$ the distance from the irradiating source. 
For the material to remain neutral, $\xi$ must be smaller than $\sim$10 (\citealt{Kallman+04}).
Adopting the 14--195 keV luminosity as $L_{\rm X} = L_{14-195~\rm{keV}} = 10^{42.36}$~erg~s$^{-1}$ (\citealt{Koss+17}) and using $r=
0.014$~pc,
we estimate the lower limit of the gas density as $\log n_{\rm xBLR}/$~cm$^{-3} > 
8$. Then, from the radial column density, the net 
thickness in the radial direction is estimated as $\Delta r_{\rm BLR}
= N_{\rm H}^{\rm BLR} / n_{\rm xBLR} < 4\times 10^{15}$~cm~$\sim
600~r_{\rm g}$,
which is 
10 times smaller than $r_{\rm BLR}$. Thus, the xBLR must be 
located in a spatially limited region if the gas distribution is smooth,
whereas it can have a larger radial extent if it is clumpy.

\begin{figure}
\centering
\includegraphics[width=0.8\linewidth]{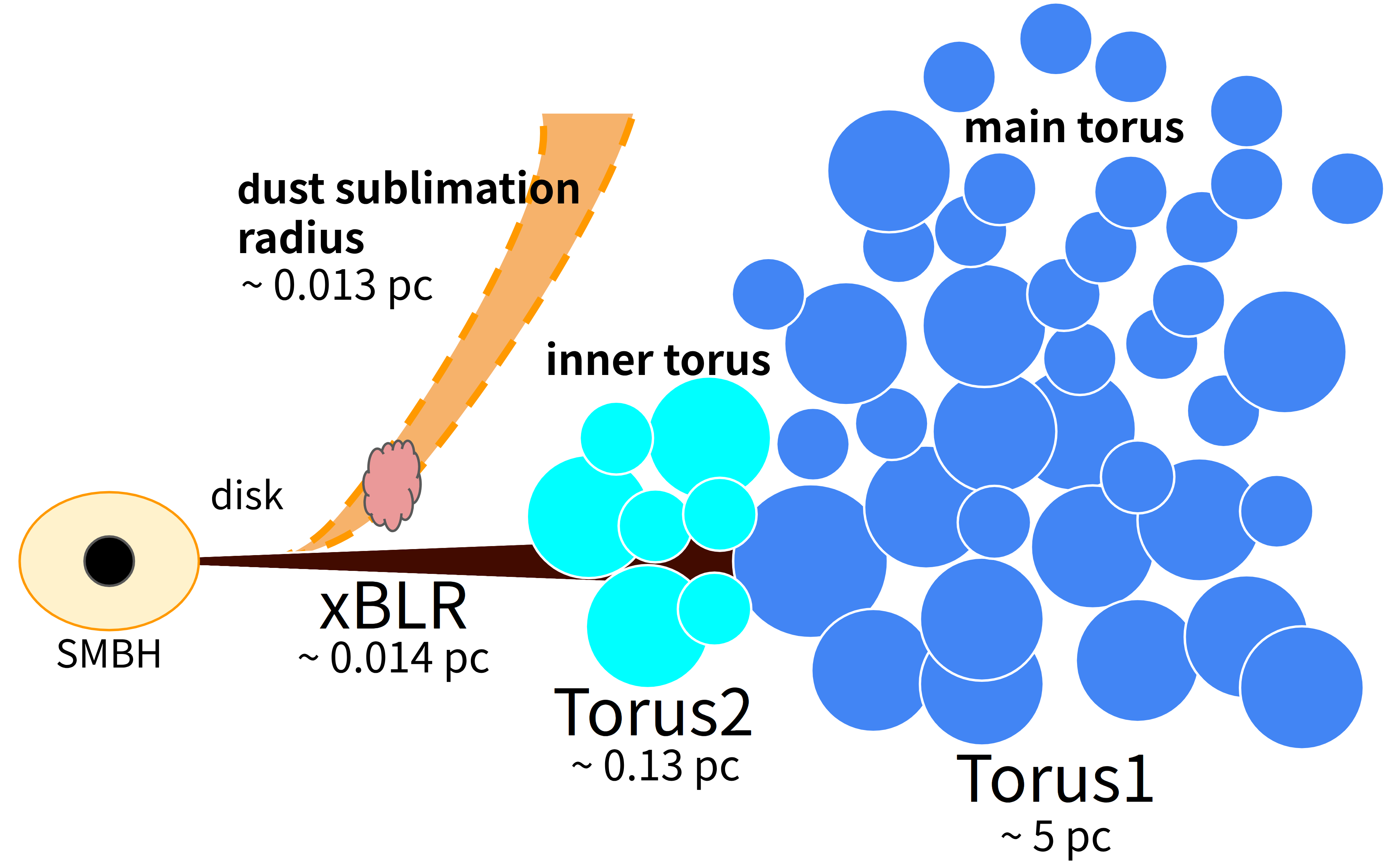}
\caption{Schematic diagram of the geometry we infer for the sub-parsec scale nuclear structure.
Around the dust sublimation radius, xBLR clouds originate or accumulate.
The ordinary torus structure consists of two components, which can be interpreted as the "inner wall" of the torus and the "mean torus body", respectively.
Estimated distances in each component are labelled in the figure, not to scale.
}
\label{fig:Ponchi}
\end{figure}

In general, the absence of broad emission lines in the scattered
(i.e., polarized) lights is explained by (1) the intrinsic absence of
the BLR (known as ``true type-2'') and/or (2) the lack of the
scattering region that allows the BLR polarized light to be observed.
Our results, confirming the presence of the xBLR in Cen~A, provide
important insights into this issue.
We propose a new explanation that the xBLR clouds in Cen~A are located
close to the inner wall of the torus ($r_{\rm torus2}/r_{\rm BLR} \sim
10$) and the geometrically thick torus ($\Delta_{\mathrm{torus}} \sim
45$~deg) obstructs light traveling from the BLR to the scattering
region. Figure~\ref{fig:Ponchi} summarizes the nuclear structure we
infer from our results including the torus and xBLR.

\subsubsection{Mass of the xBLR}\label{subsubsec:massblr}

Since X-rays are an unbiased tracer of all matter with a wide range of
physical states and are little affected by dust physics,
they give us the most complete measurement of the
total mass of the reflector, once its geometry is known. 
In our (smooth) xBLR model, the total mass of the xBLR is estimated as
\begin{equation}
M_{\rm BLR} = C_{\rm f} \frac{4 \pi}{3}  \left[ \left( r_{\rm BLR} + \Delta r_{\rm BLR}\right)^{3} - r_{\rm BLR}^{3}\right] \frac{ N_{\rm H}^{\rm BLR}}{\Delta r_{\rm BLR}} m_{\rm p} \sim 
2.0_{-1.1}^{+1.0} \ M_{\odot},
\end{equation}
where $m_{\rm p}$ is the proton mass.
The quoted uncertainty range is derived from the $\Delta C$ map in the plane of $N_{\rm H}^{\rm BLR}$ and $\sin\Delta_{\rm BLR}$ (Figure~\ref{fig:HOA-NHContour}). 
We adopt the joint 90\% confidence region of these two parameters, defined by $\Delta C \leq 4.61$ for two degrees of freedom (the yellow dashed curve in Figure~\ref{fig:HOA-NHContour}). 
The minimum and maximum values of $M_{\rm BLR}$ within this region are taken as the uncertainty range.
In this evaluation, $r_{\rm BLR}$ and $\Delta r_{\rm BLR}$ are fixed at their best-fit and derived values, respectively.
Their uncertainties are therefore not included.
Note that this value should be taken as a lower limit, because the geometry is assumed to be smooth. 
If the xBLR is clumpy and $\Delta r_{\rm BLR} \sim r_{\rm BLR}$ 
(instead of $\Delta r_{\rm BLR} \sim (
  1/10) r_{\rm BLR}$), the estimate increases by a factor of $\approx$2.
Although the mass of the BLR has been estimated in various ways, there
remain large uncertainties due to assumptions on the covering factor,
column density, and physical state of the gas, and to possible
dust extinction effects. 
\citet{Baldwin+03} estimated the mass of the BLR to be $10^{3}-10^{4}~M_{\odot}$ for luminous quasars with 1450~\AA~luminosities of 
$10^{44}$~erg~s$^{-1}$.
According to \citet{Baldwin+03} and \citet{Maiolino+10}, the mass of the BLR inferred from photoionization models scales with the AGN luminosity as 
$M_{\rm BLR}^{\rm phot} \sim 2 \times 10^{-44} (\nu L_{\nu})_{1450 {\text \AA}} M_{\odot}$.
Assuming $(\nu L_{\nu})_{1450 {\text \AA}}/L_{2-10~\rm{keV}} \sim 7$
(\citealt{Young+10, Richards+06}), this formula yields 
0.06~$M_{\odot}$ for Cen~A, which is lower than our estimate by 
more than an order of magnitude.
Note that this relation is derived from luminous quasars, and may not be directly applicable to low-luminosity AGNs like Cen~A.
In X-ray absorption studies, the mass of the BLR has been estimated from the time variability of partial covering absorbers, such as in NGC~1365 (\citealt{Maiolino+10}).
They derived the mass of the BLR to be two orders of magnitude lower than
the value inferred from photoionization models. 
This absorption method relies on the assumption that the absorbers belong to the BLR, and only captures clouds along the line of sight, leading to large uncertainties in the total number of clouds.

Assuming a BLR cloud lifetime of 100 years as derived below from the
Failed Radiatively Accelerated Dusty Outflow (FRADO) model (see below), the
xBLR formation rate can be estimated as $\dot{M}_{\rm BLR} = M_{\rm
  BLR} / 100$~yr $ 
\sim 9\textendash 30 \times 10^{-3}~M_{\odot}$~yr$^{-1}$.  The mass accretion rate onto the black
hole of Cen~A is estimated to be $\dot{M} = L_{\rm bol}/\eta c^{2} = 3 \times 10^{-3}~M_{\odot}$~yr$^{-1}$ 
(where $\eta$ is the radiative efficiency and we assume $\eta = 0.1$ for a standard disk).
This model requires that the majority ($
76\textendash91$\%) of the inflowing mass around
the dust sublimation radius be converted to xBLR clouds and retained.
These xBLR clouds are thought to be bound gravitationally, to be a
failed wind, and to provide feedback to the outer dusty torus (\citealt{Kudoh+24, Wada+26}).
This points to a robust xBLR formation mechanism that
remains efficient even in the low-Eddington regime.
We note, however, that given the low Eddington ratio of Cen~A, the innermost accretion flow may become radiatively inefficient.
In such a case, the radiative efficiency $\eta$ could be smaller than the standard
value of 0.1 assumed above (e.g., \citealt{Narayan_Yi1995, Manmoto+1997}), and the
actual mass accretion rate $\dot{M}$ could be correspondingly larger than our estimate.

The estimated mass of the xBLR and its location provide a further
constraint on how the xBLR gas is distributed. If we take the xBLR
mass to be $
0.9\textendash3~M_{\odot}$ and set a shell volume of
$V_{\rm shell} = 4\pi r_{\rm BLR}^{2} \Delta r_{\rm BLR} \sim
10^{51}$~cm$^{3}$ (here we take $\Delta r_{\rm BLR} \sim r_{\rm BLR}
\sim 0.014$~pc for the most conservative case), the smooth density
model with a filling factor $f_{\rm fill} = 1$ would give a number
density $n_{\rm smooth} = M_{\rm BLR}/(m_{\rm p} f_{\rm fill}
V_{\rm shell}) \sim (
1\textendash3)\times10^{6}$~cm$^{-3}$.
This sits one to two orders of magnitude below the lower limit of the
density estimated from the ionization parameter,
$n_{\rm xBLR} > 10^{8}$~cm$^{-3}$.
This exposes a tension between the smooth density model and the
ionization parameter constraint, suggesting that the xBLR is likely to
be clumpy rather than smoothly distributed and requiring a filling
factor of $f_{\rm fill} \sim 0.01\textendash0.03$ to reconcile the two
estimates.
In Appendix~\ref{sec:appendix} we quantify the effect of such
clumpiness on the reflection spectrum, showing that it changes the
spectral shape by $\lesssim10$\% at fixed column density, so that the
parameters derived above are unaffected, while remaining too small to
be detected with the present data.

\subsubsection{Implications for BLR Formation Models}\label{subsubsec:frado}
In this subsection, we compare our results with one of the leading BLR
formation models, called the Failed Radiatively Accelerated Dusty
Outflow (FRADO) model (\citealt{Czerny_Hryniewicz_2011, Czerny+17, Naddaf+20, Naddaf+21, Naddaf+24, Naddaf+25}). 
The FRADO model operates on the principle
of a dust-driven outflow triggered at the disk surface where
temperatures are lower than the sublimation limit ($T_{\rm disk} <
1500$~K). The dust clouds form at these cooler regions on the
accretion disk, and radiation pressure from the disk lifts these dusty
clouds off the disk surface, forming the BLR.  The dynamics of each
cloud are characterized by a recurring cycle: (1) the cloud is lifted
off the disk by radiation pressure on the dust, (2) as it is exposed
to radiation from the central SMBH corona, the dust within the cloud
sublimates, (3) deprived of the driving force provided by radiation
pressure, these clouds eventually follow a ``failed'' trajectory,
falling back to the disk surface.  This mechanism establishes a
self-consistent lifecycle for the BLR material.  The lifetime of each
BLR cloud in this cycle is estimated to be on the order of 100 years
based on nonhydrodynamical calculations.

Assuming a Shakura--Sunyaev disk (\citealt{Shakura_Sunyaev_1973}), the dust formation radius, where the disk temperature drops below the dust sublimation temperature ($T_{\rm disk} = 1500$~K), is estimated to be around 1000~$r_{\rm g}$.
At the projected location of the xBLR ($
5900~r_{\rm g}$), the disk surface temperature is lower, estimated at $T_{\rm disk}(
5900~r_{\rm g}) \sim 400$~K.
Thus, the xBLR is located outside the dust formation radius.
Given this separation, while dust may play a role in cloud formation and inflate the BLR, it is necessary to consider additional physical processes to transport the xBLR clouds outward and accumulate them at such a specific location.

For a black-hole mass of $10^{8}~M_{\odot}$ and an Eddington ratio of 0.01, 
\citet{Naddaf+20} calculated a BLR opening angle ratio of approximately 0.03, which corresponds to an opening angle of $ \sim 2$~deg.
In contrast, we find an xBLR opening angle of $> 6$~deg.
This larger angle aligns with the findings of \citet{Naddaf+20} for high-Eddington-ratio ($\sim 1$) AGNs.
As noted by \citet{Czerny+17}, the insufficiency of radiation pressure in standard dust-driven winds implies the presence of optically thick clouds or additional mechanisms.
Our results provide the first X-ray-based constraints that can be used to further refine and update the standard BLR formation scenario.

The inflow required by the xBLR formation rate estimated above is indeed
observed on larger scales.
\citet{Evangelista+26} found that shocks excite the H$_2$ gas,
regulate its temperature, and prevent cooling below 100~K.
These shocks may drive angular momentum loss and help fuel the nucleus.
As evidence for such an inflow, they identified a coherent, low-dispersion
(70~km~s$^{-1}$) streamer spiralling inward in the central
$170 \times 100$~pc$^{2}$ region.
\citet{Jones+26} further showed that the warped central dust disk of Cen~A
was formed in a gas-rich merger, and that star formation there is primarily
regulated by merger-accreted gas rather than by the jet.
These results provide the large-scale counterpart of the mass budget
required by our estimate.

\subsubsection{SMBH Mass Estimation Using High Energy-Resolution X-ray Spectra}\label{subsubsec:massestimation}

The detection of this hidden xBLR has an impact extending well beyond Cen~A,
providing a powerful new tool to estimate SMBH masses in obscured AGNs. 
In this section, we discuss this method, enabled by \textit{XRISM}'s capability to
spectroscopically resolve broad X-ray line components. While standard
techniques using the line width of the (optical) BLR and continuum
luminosity in the UV to optical bands have been successful in estimating
their SMBH masses, these cannot be applied to type 2 AGNs. Thanks to the
strong penetrating power of X-rays, a BLR hidden in the optical band 
can still be observed in X-rays, as demonstrated in our
observations. Since a major fraction of AGNs in the universe are
obscured (\citealt{Ueda+14}), establishing a method of SMBH mass estimation for
obscured AGNs is crucial to understanding the growth history of whole
SMBHs.

Our purpose is to extend the established relation in the optical band
using the FWHM of the H$\beta$ line and $L_{{5100}\text{\AA}}$ (e.g.,
\citealt{Bentz+09, Trakhtenbrot_Netzer_2012}) to the X-ray regime. Here we assume that the optical BLR is a part
of the xBLR and their locations are the same. In fact,
quasi-simultaneous observations of the Seyfert 1.5 galaxy NGC~4151
with \textit{XRISM} and \textit{Seimei} telescope have shown that the
widths of the H$\beta$ line and the Fe K$\alpha$ line are similar to
each other (Noda et al., in prep.), supporting this interpretation. 
We can estimate the intrinsic luminosity at 5100~\AA~of an obscured AGN
from the 14--195 keV luminosity by the relation
\begin{eqnarray}
  \log L_{5100\text{\AA}} &=& \log L_{14\textendash195\rm{keV}} - 0.32~(\pm 0.50), \label{eq:L5100_bat}
\end{eqnarray}
where the luminosities are in units of ${\rm erg~s^{-1}}$ and the
value in parentheses is the $1\sigma$ scatters in dex.
We have derived this relation from 327 broad-line AGNs of the BAT AGN
Spectroscopic Survey (BASS) utilizing the most up-to-date estimates of
$L_{5100\text{\AA}}$ (see Appendix~\ref{sec:appendix_Lx_L5100}).

The SMBH mass estimation formula based on optical reverberation
mapping is (\citealt{Trakhtenbrot_Netzer_2012})
\begin{equation}\label{eq:mbh_hb}
  \log \frac{M_{\rm BH}(\mathrm{H\beta})}{M_{\odot}} = 8.02
+ 0.65 \log \left( \frac{L_{5100\text{\AA}}}{10^{46}~{\rm erg~s^{-1}}} \right)
+ 2 \log \left[ \frac{{\rm FWHM}(\rm{H\beta})}{10^{3}~{\rm km~s^{-1}}} \right],
\end{equation}
where $8.02 = \log (1.05 \times 10^{8})$.
This relation assumes a virial factor of $f = 1$.
The index of $0.65$ is that of the radius-luminosity ($R$--$L$) relation of the H$\beta$ BLR. 
We generalize Equation~(\ref{eq:mbh_hb}) for the Fe~K$\alpha$ line
and hard X-ray luminosity as
\begin{equation}\label{eq:mbh_feka}
\log \frac{M_{\rm BH}(\mathrm{Fe~K\alpha})}{M_{\odot}} = 8.02 + \log A
+ B \left[ \log \left( \frac{L_{14\textendash195\rm{keV}}}{10^{46}~{\rm erg~s^{-1}}} \right) - 0.32 \right]
+ 2 \log \left[ \frac{{\rm FWHM}(\rm{Fe~K\alpha})}{10^{3}~{\rm km~s^{-1}}} \right].
\end{equation}
Here $A$ is a dimensionless coefficient that scales the normalization of the H$\beta$ relation, $1.05\times10^{8}$, and
$B$ is the power-law index of the luminosity dependence.
Both must be calibrated for the Fe~K$\alpha$ line.
Empirically, the index of the optical $R$--$L$ relations lies in the range of $0.5$--$0.7$ (\citealt{Trakhtenbrot_Netzer_2012}), 
which can differ among emission lines (e.g., Mg~II and H$\beta$).

We first apply Equation~(\ref{eq:mbh_feka}) to Cen~A with the optical values, $A = 1$ and $B = 0.65$.
We use ${\rm FWHM}(\mathrm{Fe~K\alpha}) = 6000~{\rm km~s^{-1}}$ and $\log L_{14\textendash195\rm{keV}}/{\rm erg~s^{-1}} = 42.36$ (\citealt{Koss+17}).
This gives $\log L_{5100\text{\AA}}/{\rm erg~s^{-1}} = 42.04$ and $\log M_{\rm BH}(\mathrm{Fe~K\alpha}) / M_{\odot} = 7.00\pm0.34$.
The uncertainty is quoted at $1\sigma$, combined in quadrature from two independent sources:
the statistical error in the Fe~K$\alpha$ line width and 
the scatter in the conversion to $L_{5100\text{\AA}}$ multiplied by $B$.
The best estimate is lower than the mass adopted in Section~\ref{sec:intro} ($\log M_{\rm BH}/M_{\odot} = 7.7$) and 
the difference is $0.70$~dex (a factor of about 5).
For $B = 0.5$ with $A = 1$, 
we obtain $\log M_{\rm BH}(\mathrm{Fe~K\alpha})/M_{\odot} = 7.60\pm0.26$.
This is consistent with $7.7$ within $1\sigma$.
These results suggest that this method of SMBH mass estimation using the broad Fe~K$\alpha$ line is promising.

To make the relation exactly consistent with the adopted mass, 
$A$ and $B$ should satisfy $\log A - 3.96\,B = -1.88$ (e.g., $A\approx1.3$ for $B=0.5$).
An earlier study reported a larger mass of $\log M_{\rm BH}/M_{\odot} = 8.38$ (\citealt{Silge+05}; see Section~\ref{sec:intro}).
For this mass, $B=0.5$ and $A\approx6$ would be required.
Determining these coefficients requires a sample of AGNs with broad Fe~K$\alpha$ widths and independent mass measurements covering a wide range of X-ray luminosity.

However, this method is still in its early stages and several caveats remain. 
First, adopting the optical values for $A$ and $B$ relies on the assumption that the xBLR
is co-spatial with the optical BLR, a hypothesis that remains to be confirmed (\citealt{Marin+25}). 
This needs to be addressed through simultaneous observations of xBLR with \textit{XRISM} 
and optical BLR using a large sample of unobscured AGNs.

Second, geometry-induced bias may exist.  The optical R-L relations
are biased toward face-on systems; thus, applying them to edge-on
systems (obscured AGNs) may introduce systematic errors due to the
anisotropic velocity structure of the xBLR (see discussion in
\citet{Fujiwara+26} for NGC~4388).
It is therefore important to perform reverberation mapping of Fe
K$\alpha$ (e.g., \citealt{Miller+26}) to directly determine 
the relation for obscured AGNs. 
Third, the conversion from the 14--195~keV luminosity to the
5100~\AA~luminosity has a $1\sigma$ scatter of $0.50$~dex (Appendix~\ref{sec:appendix_Lx_L5100}).
This is the dominant term in the uncertainty of our estimate.
This could be improved by considering the dependence of the
spectral energy distribution on AGN luminosity and/or Eddington ratio.

A black-hole mass estimation method for obscured AGNs was recently proposed by \citet{LaMassa+25}. 
Their method is based on the virial relation of \citet{Baron_Menard_2019}. 
Therefore, the basic framework is similar to ours, but the method to derive each parameter is different. 
Instead of measuring the velocity width kinematically, 
\citet{LaMassa+25} adopt an empirical estimate from the optical narrow-line ratio $L_{\rm [O\,III]}/L_{\rm narrow~H\beta}$. 
Our method has the advantage of directly measuring the velocity width from the Fe K$\alpha$ line. 
\citet{LaMassa+25} report a luminosity scatter of $0.68$~dex in $L_{5100\text{\AA}}$ derived from the intrinsic 2--10~keV luminosity. 
This is comparable to the $0.52$~dex scatter that we obtain for the same band (Appendix~\ref{sec:appendix_Lx_L5100}). 
Our adopted conversion from the 14--195~keV luminosity has a scatter of $0.50$~dex.
However, a direct comparison for Cen~A is difficult 
because the strong nuclear extinction makes it hard to reliably measure the narrow-line ratio, 
and it does not satisfy their selection criteria. 
Applying both methods to a common sample of nearby obscured AGNs with high-resolution X-ray data is left as future work.

Recent \textit{JWST}/MIRI observations have characterized the circumnuclear region of Cen~A in detail.
\citet{Alonso-Herrero+25} studied the kinematics of the ionized gas and revealed a fast outflow along the radio jet.
\citet{Evangelista+26} traced the feeding and feedback of the warm molecular gas,
as described in Section~\ref{subsubsec:frado}.
\citet{Jones+26} characterized the embedded star formation in the warped dust disk.
None of these studies, however, constrains the torus or the BLR.
A direct test of the co-spatiality would be near-infrared spectroscopic observations of the core, 
targeting hydrogen recombination lines such as Pa$\beta$ and Br$\alpha$ (e.g., with \textit{JWST}/NIRSpec). 
Existing near-infrared spectroscopy shows no broad component in Pa$\alpha$ or Br$\gamma$ (\citealt{Goold+26}), 
which is consistent with the strong extinction toward the nucleus. 
A search in the longer-wavelength lines, where the extinction is smaller, is therefore the natural next step.

\subsection{Chemical Abundances}\label{subsec:abundanceratio}

\begin{figure}
  \centering
  \includegraphics[width=0.8\linewidth]{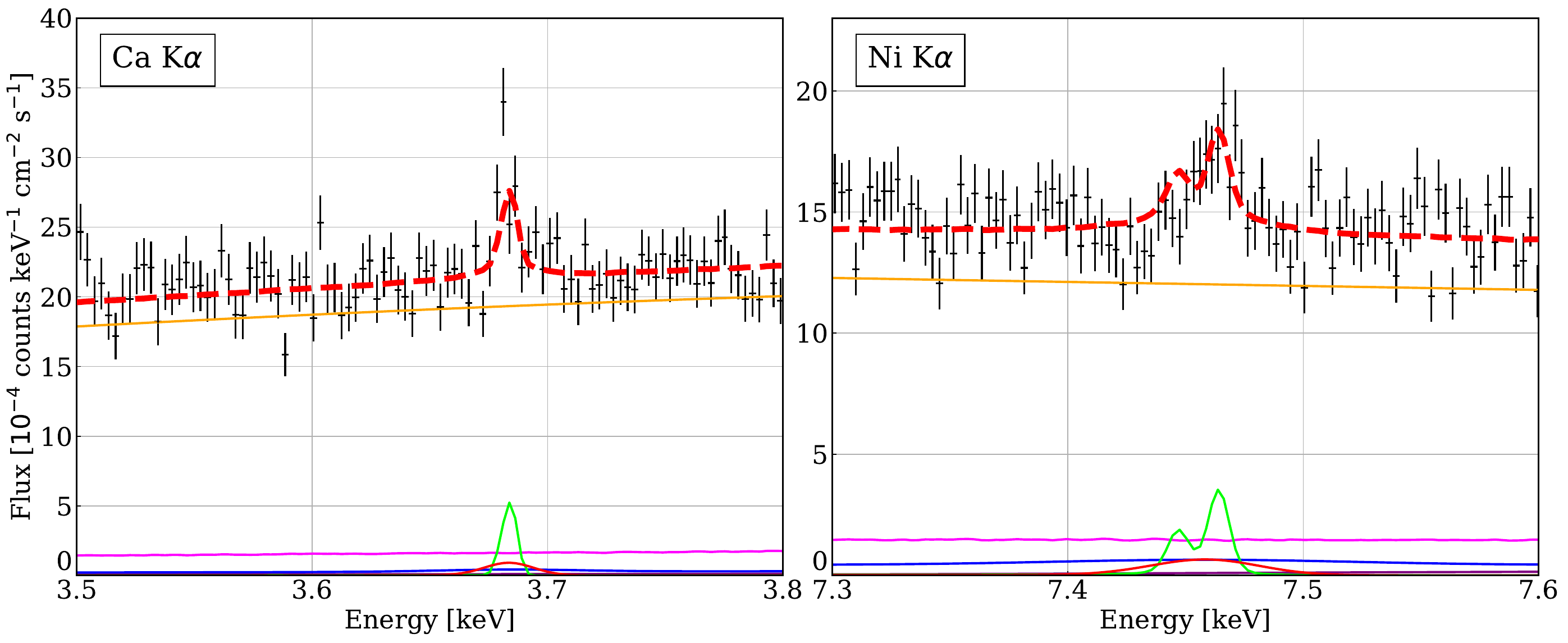}
  \caption{Zoom-in around the fluorescence lines, Ca K$\alpha$ and Ni K$\alpha$. 
    The \textit{XRISM}/Resolve data points and the best-fit model are shown in the same way as in the right panel of Figure~\ref{fig:FeZoom}.
  }
  \label{fig:CaNiZoom}
\end{figure}

In this subsection, we briefly discuss the abundance ratios of Ca and
Ni to Fe in the torus and xBLR of Cen~A, based on the spectral fitting
results using the XCLUMPY and xBLR reflection models.  The high-energy
resolution spectroscopy with \textit{XRISM}/Resolve allows us to
detect and precisely measure the intensities of these fluorescent
lines for the first time in a radio galaxy.
Figure~\ref{fig:CaNiZoom} shows the zoomed-in spectra around the Ca
K$\alpha$ (3.69 keV in rest frame) and Ni K$\alpha$ (7.48 keV)
lines. In our best-fit model,
the abundance ratios of Ca and Ni to Fe are found to be $A_{\rm
  Ca}/A_{\rm Fe} = 
1.10_{-0.11}^{+0.20}$ and $A_{\rm Ni}/A_{\rm Fe} =
1.35_{-0.18}^{+0.07}$.

As an independent check, we estimated $A_{\rm Ni}/A_{\rm Fe}$ from the 
observed Fe K$\alpha$-to-Ni K$\alpha$ photon flux ratio 
($I_{\rm Fe~K\alpha}/I_{\rm Ni~K\alpha} \approx 12$) using the empirical 
relation of \citet{Yaqoob_Murphy_2011}. In the optically thin limit, 
this relation is independent of the continuum shape and the covering factor. 
When we adjust their normalization to the \citet{Lodders+09} solar abundance 
scale adopted in this paper, we obtain $A_{\rm Ni}/A_{\rm Fe} \approx 1.6$. 
This value is qualitatively consistent with our best-fit value. 
However, we do not expect exact quantitative agreement because 
the relation of \citet{Yaqoob_Murphy_2011} strictly holds only for 
$N_{\rm H}\lesssim2\times10^{22}$~cm$^{-2}$. This limit is lower than 
the column density inferred for the reflecting material in Cen~A.

However, we note some limitations in our analysis that may affect the
accuracy of these abundance ratio measurements besides the statistical
uncertainties. The data taken with \textit{NuSTAR} do not show
strong evidence of a Compton hump in the hard X-ray band, offering us
only upper limits on the reflection continuum strength.  It leads to
uncertainties
in the ratio of fluorescent line intensities to that of the reflection
continuum, which is a key quantity that determines the elemental
abundance relative to hydrogen. Our current estimate of the reflection
continuum strength depends on the iron to hydrogen abundance
ratio, which is assumed to be the solar value. In addition,
the uncertainty on the reflection continuum weakens the
constraints on the torus geometric structure and affects the
evaluation of self-absorption. It becomes more important for modeling
of the reflection continuum at soft energies, thereby impacting the
evaluation of the abundance ratio of $A_{\rm Ca}/A_{\rm Fe}$.

Bearing these caveats in mind, the abundance ratio patterns we
obtained for the reflecting materials in Cen~A provide insights
into the possible origin of the mass reservoirs around the AGN. For
the Circinus Galaxy, \citet{XRISM_Circinus2026} reported that the
abundance pattern of the reflecting materials shows sub-solar $A_{\rm
  Ar}/A_{\rm Fe}$ and $A_{\rm Ca}/A_{\rm Fe}$ ratios, while the
$A_{\rm Ni}/A_{\rm Fe}$ ratio is super-solar. They concluded that the
central SMBH is fed by gas enriched predominantly by recent
core-collapse supernovae from progenitors less massive than $20~M_{\odot}$, reflecting the active nuclear star formation in the
spiral host. In contrast, our results for Cen~A show solar value for
$A_{\rm Ca}/A_{\rm Fe}$ and nominally super-solar value for $A_{\rm Ni}/A_{\rm
  Fe}$, implying a different enrichment history. Given the limited
number of elements and large statistical and systematic uncertainties,
we leave detailed modeling of the abundance pattern with supernova
nucleosynthesis models for future work. Future deeper observations
that enable us to better determine the abundance ratios including
other elements (particularly Ar and Mn) would be required to exploit
the abundance-pattern diagnostics.

\section{Conclusions}\label{sec:conclusion}

We summarize our main findings from the broadband spectral analysis of
Cen~A using simultaneous \textit{XRISM} (Resolve and Xtend) and \textit{NuSTAR}
observations. For our analysis, we developed radiative transfer-based
numerical models representing the reflection continuum and line
components from the torus and the xBLR, employing the SKIRT framework.

\begin{enumerate}
  
\item The velocity profile of the Fe K$\alpha$ line is well reproduced with three Gaussian kernels, similarly to the case of NGC 4388. 
This feature may be common in AGNs when sufficient photon statistics of Fe K$\alpha$ are obtained. 
We interpret that the two narrower components (with FWHMs of $
1900\pm400$~km~s$^{-1}$ and $
310\pm50$~km~s$^{-1}$) are from the inner and main torus regions,
 whereas the broad one ($
 6000_{-800}^{+1100}$~km~s$^{-1}$) is from the xBLR. 
 This enables physical decomposition of the broadband spectra, consisting of the primary emission, ionized emission, and reflection components from the torus and xBLR. 
 We demonstrate that high-energy-resolution X-ray spectroscopy can detect a deeply buried BLR that remains undetected in
 the optical-to-infrared bands due to severe extinction ($A_V \approx$ 22--55~mag).

\item Our results confirm a picture that the torus in Cen~A has a large covering factor ($\Delta_{\rm torus} = 
44_{-4}^{+3}$~deg) 
with moderate hydrogen column densities ($N_{\rm H}^{\rm eq} = 
33_{-3}^{+6} \times 10^{22}$~cm$^{-2}$). 
This ``buried'' nature  may explain the absence of broad emission lines even in the polarized optical spectra.

\item The representative distances of the inner and main torus regions from the SMBH  are $
0.13_{-0.04}^{+0.07}$~pc and $
5_{-1}^{+2}$~pc, respectively. 
This is about 10 times larger than the estimated dust sublimation radius, 
implying that the innermost radius of the torus is not determined by dust-related physics in a low-luminosity AGN.

\item We suggest that the xBLR has a substantial covering factor $0.11-0.36$ with a moderate hydrogen column density (
  $5_{-3}^{+5} \times10^{23}$~cm$^{-2}$) and is located at 0.014~pc from the SMBH. 
The estimated total mass in the xBLR is $>
0.9~M_{\odot}$, which is larger than those in other AGNs based on Balmer lines or X-ray absorption variability. 
The xBLR mass relative to the accretion rate suggests a robust xBLR formation mechanism that remains highly efficient even in low-Eddington ratio regimes, providing a new constraint on BLR production mechanisms.

\item We propose a novel method to estimate the SMBH mass of obscured AGNs from the xBLR line width and the X-ray luminosity applicable to obscured AGNs. 
The X-ray luminosity is converted to $L_{5100\text{\AA}}$ with a relation derived from BASS broad-line AGNs.
With the constants of the optical H$\beta$ relation, 
this method yields $\log M_{\rm BH}/M_{\odot} = 7.00\pm0.34$ from the observed 14--195~keV luminosity.
This is smaller than the kinematic mass by a factor of about 5.
With a flatter luminosity dependence ($B=0.5$), the method yields $7.60\pm0.26$.
This is consistent with the kinematic mass within $1\sigma$.
Determining the two constants requires a sample of AGNs with independent mass measurements.

\item We derived abundance ratios of $A_{\rm Ca}/A_{\rm Fe} \sim 1.1$, and $A_{\rm Ni}/A_{\rm Fe} \sim 1.4$ in the reflecting material of Cen~A.

\end{enumerate} 

\section{Software and third party data repository citations}\label{sec:cite}

\facilities{\textit{XRISM}, \textit{NuSTAR}}
\software{XSPEC (\citealt{Arnaud_1996}), SKIRT (\citealt{Baes+03, VanderMeulen+23})}

\appendix
\section{Effects of Clumpiness in xBLR Gas on the Model Spectra}\label{sec:appendix}

The xBLR model in the main text assumes a smooth geometry, yet observational and theoretical studies suggest that the BLR is not smooth but rather inhomogeneous.
If the BLR has a clumpy density distribution, the self-absorption within individual clouds can differ from that of a smooth model with the same average column density along the line of sight.
This difference could alter the ratio of the Fe K$\alpha$ line intensity to the Compton hump intensity, thereby introducing a systematic error into the abundance ratio measurements derived in the main text.
This appendix quantifies the systematic impact of BLR clumpiness on the X-ray reflection spectrum using SKIRT simulations and evaluates its implications for the main-text results and conclusions.
We also assess whether the high-energy-resolution data from \textit{XRISM}/Resolve can place quantitative constraints on clumpiness.

For the simulations, we adopt the same xBLR geometry as in the main analysis, but replace the smooth density distribution with a clumpy one composed of spherical clumps randomly distributed with a uniform probability density within the BLR volume.
We parameterize clumpiness by varying the number of clumps along the line of sight, $N_{\rm clump,LOS} = 5, 10, 20, 50$.
At $N_{\rm clump,LOS} = 5$ the model is most clumpy; 
increasing $N_{\rm clump,LOS}$ reduces individual clump sizes and column densities, and the model converges toward a smooth distribution.
We assume that each clump has a radius of $R_{\rm clump} = r_{\rm min}/N_{\rm clump,LOS}$.
In this setting, the volume filling factor (the total clump volume relative to the BLR volume) remains constant across all models.
In order to maintain a constant $N_{\rm H,LOS}$ across all models, the hydrogen number density of each clump is adjusted accordingly.
The computational grid is refined so that each clump is resolved by a minimum of $3\times3\times3$ cells.
All models share the following fixed parameters: $r_{\rm min}=0.01$~pc, $r_{\rm max}= 10\times r_{\rm min}=0.1$~pc, $N_{\rm H,LOS}=10^{23}$~cm$^{-2}$, $\Delta_{\rm BLR}=15$~deg, $\Gamma=1.8$, $E_{\rm cut}=370$~keV, and solar abundances.
We compare these clumpy models with the smooth model used in the main analysis in terms of their reflection spectra.

\begin{figure*}[tbp]
\centering
\includegraphics[width=0.8\textwidth]{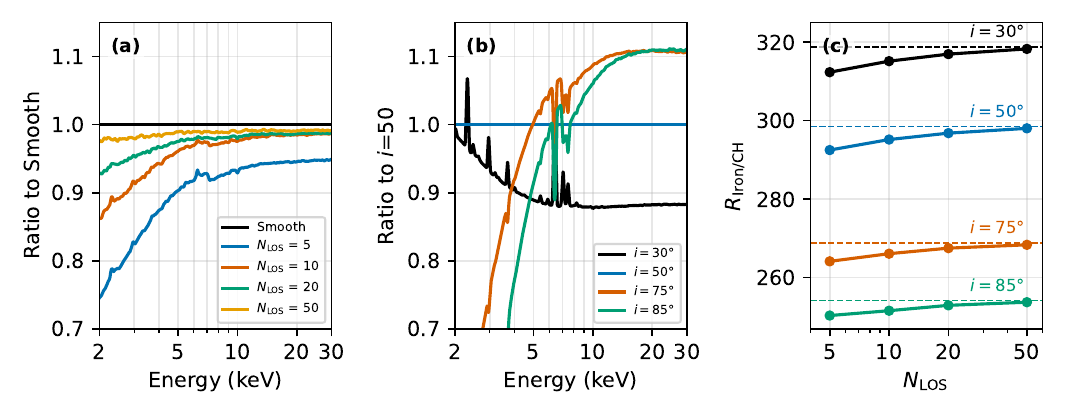}
\caption{
Dependence of the spectral shape produced by a clumpy reflector on the line-of-sight clump number $N_{\rm LOS}$ and the inclination angle $i$.
All spectra shown here are smoothed with a Gaussian of 50~eV (at 6~keV) for display purposes.
(a) Ratio of the reflection spectrum (reflection continuum plus line components) at $i = 50^{\circ}$ relative to the smooth reflector (Smooth), shown over the 2--30 keV energy range.
The blue, red, green, and orange lines represent the ratios for the clump models with $N_{\rm LOS} = 5, 10, 20, 50$, respectively.
The black line is for the reference Smooth model.
(b) Spectra of the smooth reflector model at $i = 30^{\circ}, 50^{\circ}, 75^{\circ}, 85^{\circ}$ normalized by the $i = 50^{\circ}$ spectrum.
The black, blue, red, and green  lines represent the ratios of the Smooth model at $i = 30^{\circ}, 50^{\circ}, 75^{\circ}$, and $85^{\circ}$, respectively.
(c) Dependence of $R_{\rm Iron/CH}$ on $N_{\rm LOS}$. 
Here, $R_{\rm Iron/CH}$ is defined as the ratio between the iron-line intensity (6.0--6.8 keV) and the Compton-hump intensity (15--40 keV).
Solid lines with circles represent the clump models for each inclination, and dashed lines represent the Smooth model at the same inclination.
The black, blue, red, and green colors indicate $i = 30^{\circ}, 50^{\circ}, 75^{\circ},$ and $85^{\circ}$, respectively.
}
\label{fig:clumpiness}
\end{figure*} 

The ratios to the smooth reference model are presented in Figure~\ref{fig:clumpiness}(a).
Spectral differences between the clumpy and smooth models are  
$\sim 5\textendash20\%$ at $N_{\rm clump,LOS}=5$,
$\sim10\%$ at $N_{\rm clump,LOS}=10$, 
and below $2\%$ at $N_{\rm clump,LOS}=50$.
In order to quantify the spectral differences, we focus on the flux ratio $R_{\rm Iron/CH}$ of the Fe~K$\alpha$ line (6.0--6.8~keV) to the Compton hump (15--40~keV). 
As $N_{\rm clump,LOS}$ increases, $R_{\rm Iron/CH}$ increases and converges toward the smooth reference value from below (Figure~\ref{fig:clumpiness}(c)).
This trend can be understood as follows: 
in low $N_{\rm clump,LOS}$ models, each clump has a large column density and enhances self-absorption, leading to a more suppressed Fe~K$\alpha$ line intensity, while high energy photons of the Compton hump are less susceptible to self-absorption.
To assess the relative importance of clumpiness, we compare this effect with the spectral variation caused by changes in the inclination angle.
Figure~\ref{fig:clumpiness}(b) shows the spectral variation computed for each model with inclination angles of $30$, $50$, $75$, and $85$~deg.
A change in inclination angle alters the $z$-direction column density and hence the degree of self-absorption, producing changes of the spectral shape, mainly on hardness ratios.
This variation exceeds the $\lesssim10\%$ effect due to clumpiness, so the systematic error arising from clumpiness is negligible.
Consequently, quantitative constraints on clumpiness remain beyond the reach of current observations, even with the \textit{XRISM}/Resolve data.

\section{Relation between the 5100~\AA\ and X-ray Luminosities}\label{sec:appendix_Lx_L5100}

The relation between the 14--195~keV and $L_{5100\text{\AA}}$
luminosities was reported by \citet{Koss+17} based on the BAT AGN
Spectroscopic Survey Data Release 1. Here we rederive the relation by
utilizing more recent results of $L_{5100\text{\AA}}$ (\citealt{Mejia-Restrepo+22}) based on a
larger sample of broad-H$\beta$ AGNs in BASS DR2 (\citealt{Koss+22}). The
final sample consists of 327 AGNs after excluding 34 blazars and 3
outliers\footnote{Two of them are BATIDs 143 and 234, which show anomalous
  $L_{5100\text{\AA}}$ to $L_{14-195\rm{keV}}$ ratios. 
  The other is BATID 277, for which \citet{Ricci+17BASS} and \citet{Koss+22} appear to adopt different counterparts.}. 
  The observed rest-frame 14--195~keV
luminosities (not corrected for absorption) are taken from Table~13 of
\citet{Ricci+17BASS}.

\begin{figure*}
\centering
\includegraphics[width=0.5\linewidth]{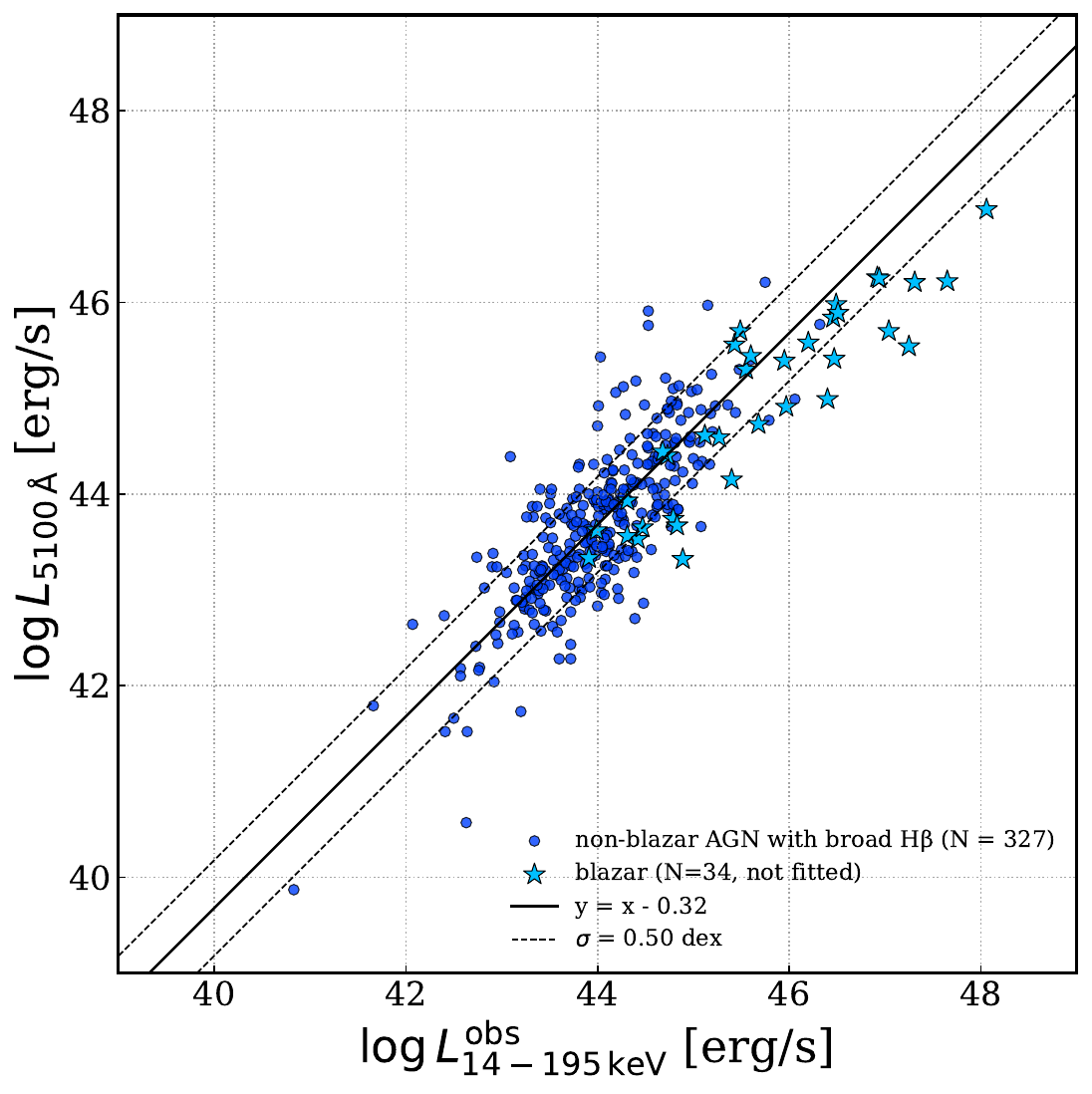}
\caption{Relation between $L_{5100\text{\AA}}$ and 14--195 keV
  luminosity for the BASS broad-line AGNs (blue circles). Blazars,
  which are not included in the fit, are marked in light-blue stars. 
  The solid line is the best-fit relation $y = x + a$, where $x$ is the logarithm of
  the X-ray luminosity and $y = \log L_{5100\text{\AA}}$. The dashed
  lines show the region within $\pm1\sigma$ scatter.}
\label{fig:L5100_Xray}
\end{figure*}

Figure~\ref{fig:L5100_Xray} plots the relation between the $L_{5100\text{\AA}}$ and $L_{14-195\rm{keV}}$ of this
sample. We fit these data with $y = x + a$, where $x$ and $y$ are the
logarithms of the 14--195 keV and 5100~\AA\ luminosities, respectively.
We obtain $a = -0.32\pm 0.03$ with a 1$\sigma$ scatter of $0.50$~dex,
which we adopt in Equation~(\ref{eq:L5100_bat})\footnote{If we use the
  2--10 keV intrinsic luminosities in Table~13 of \citet{Ricci+17BASS},
  instead of the observed 14--195 keV ones, we obtain $a=+0.11\pm0.03$ with a
  1$\sigma$ scatter of $0.52$~dex.}.
There are caveats, however. 
First, the optical and X-ray data are not simultaneous and could be affected
by time variability. Second, the $L_{5100\text{\AA}}$ values in
\citet{Mejia-Restrepo+22} include the host-galaxy light, which may
therefore make $L_{5100\text{\AA}}$ overestimated, especially at low
luminosities.

\begin{acknowledgments}
Y.N. acknowledges financial support from Grant-in-Aid for the Japan Society for the Promotion of Science (JSPS) Fellows (24KJ1507).
This work was partially supported by JSPS KAKENHI Grant Numbers JP25H00671 (Y.U.), JP26K17197, JP21K13958 (M.M.), and JP25KJ1868 (T.K.).
D.B. acknowledges financial support from the Centre national d'\'etudes spatiales (CNES), France (ROR: \href{https://ror.org/04h1h0y33}{https://ror.org/04h1h0y33}) within the framework of the XMM-Newton mission.
L.G. acknowledges support from the Canadian Space Agency grant 25EXPRSM1 and 18XARMSTMA.
T.Y. acknowledges support from CRESST II under NASA award number 80GSFC24M0006.
B.V. acknowledges support through the European Space Agency (ESA) Research Fellowship in Space Science. 
M.M. acknowledges support from Yamada Science Foundation.
\end{acknowledgments}

\clearpage

\bibliography{main}{}
\bibliographystyle{aasjournalv7}

\end{document}